# Direct images and spectroscopy of a giant protoplanet driving spiral arms in MWC 758

Kevin Wagner,[1,*,**] Jordan Stone,[2] Andrew Skemer,[3] Steve Ertel,[1] Ruobing Dong,[4] Dániel Apai,[1,5] Eckhart Spalding,[6] Jarron Leisenring,[1] Michael Sitko,[7] Kaitlin Kratter,[1] Travis Barman,[5] Mark Marley,[5] Brittany Miles,[1] Anthony Boccaletti,[8] Korash Assani,[9] Ammar Bayyari,[10] Taichi Uyama,[11,12,13] Charles E. Woodward,[14] Phil Hinz,[3] Zackery Briesemeister,[3] Kellen Lawson,[15] François Ménard,[16] Eric Pantin,[17] Ray W. Russell,[18] Michael Skrutskie,[9] and John Wisniewski[19]

[1] Department of Astronomy and Steward Observatory, University of Arizona, 933 N Cherry Ave, Tucson, AZ 85721 USA
[2] Naval Research Laboratory, 4555 Overlook Ave SW, Washington, DC 20375 USA
[3] Department of Astronomy & Astrophysics, University of California Santa Cruz, 1156 High Street, Santa Cruz, CA 95064 USA
[4] Department of Physics and Astronomy, University of Victoria, 3800 Finnerty Road, Victoria, BC V8P 5C2 Canada
[5] Lunar and Planetary Laboratory & Dept. of Planetary Sciences, University of Arizona, 1629 E University Blvd, Tucson, AZ 85721 USA
[6] The University of Sydney, NSW 2006, Australia
[7] Space Science Institute, 4765 Walnut St STE B, Boulder, CO 80301 USA
[8] Observatoire de Paris-PSL, CNRS, 5 place Jules Janssen, 92195 Meudon, France
[9] Department of Astronomy, University of Virginia, 530 McCormick Rd, Charlottesville, VA 22904 USA
[10] Department of Physics and Astronomy, University of Hawaii, 2505 Correa Road Watanabe 416. Honolulu, Hawaii 96822 USA
[11] Infrared Processing and Analysis Center, California Institute of Technology, 1200 E. California Blvd., Pasadena, CA 91125, USA
[12] NASA Exoplanet Science Institute, Pasadena, CA 91125, USA
[13] National Astronomical Observatory of Japan, 2-21-1 Osawa, Mitaka, Tokyo 181-8588, Japan
[14] Minnesota Institute for Astrophysics, University of Minnesota, 116 Church Street, SE, Minneapolis, MN 55455 USA
[15] Homer L. Dodge Department of Physics and Astronomy, The University of Oklahoma 440 W. Brooks St. Norman, OK 73019
[16] Univ. Grenoble Alpes, CNRS, IPAG, F-38000 Grenoble, France
[17] Département d'Astrophysique, CE Saclay, Orme des Merisiers, Bât 709, 91191 Gif-sur-Yvette, France
[18] The Aerospace Corporation, 2310 E. El Segundo Blvd. El Segundo, CA 90245 USA
[19] NASA Headquarters, 300 Hidden Figures Way SW, Washington, DC 20546, USA
*   NASA Hubble Fellowship Program – Sagan Fellow
** correspondence to kevinwagner@arizona.edu

**Understanding the driving forces behind spiral arms in protoplanetary disks remains a challenge due to the faintness of young giant planets. MWC 758 hosts such a protoplanetary disk with a two-armed spiral pattern that is suggested to be driven by an external giant planet. We present new thermal infrared observations that are uniquely sensitive to redder (i.e., colder or more attenuated) planets than past observations at shorter wavelengths. We detect a giant protoplanet, MWC 758c, at a projected separation of ~100 au from the star. The spectrum of MWC 758c is distinct from the rest of the disk and consistent with emission from a planetary atmosphere with $T_{\text{eff}} = 500 \pm 100$ K for a low level of extinction ($A_V$≤30), or a hotter object with a higher level of extinction. Both scenarios are commensurate with the predicted properties of the companion responsible for driving the spiral arms. MWC 758c provides evidence that spiral arms in protoplanetary disks can be caused by cold giant planets or by those whose optical emission is highly attenuated. MWC 758c stands out both as one of the youngest giant planets known, and also as one of the coldest and/or most attenuated. Furthermore, MWC 758c is among the first planets to be observed within a system hosting a protoplanetary disk.**

Giant protoplanets interact gravitationally with their birth disks, driving gaps and large-scale spiral structures that alter the environment for subsequent planet formation (*1, 2*). Several protoplanetary disks that are stable to their own self-gravity (i.e., that should not show spiral structure due to instabilities) have been observed with global spiral morphologies that resemble predictions from companion-disk interaction models (e.g., *3, 4*). However, this picture has only been confirmed for a handful of systems with stellar and brown dwarf companions (*2, 5, 6*). Several studies have hypothesized that the spiral arms in the other disks are caused by giant planets that formed with lower initial entropy (*7, 8*) or by those whose optical emission is heavily attenuated by circumstellar or circumplanetary material. If most protoplanets form cold or significantly reddened, this would explain the lack of previously detected planets in systems with spiral arms, and more broadly the low observed yield of many past searches for protoplanets (e.g., *9, 10*). Around one such spiral disk, MWC 758, thermal infrared observations have revealed a very red candidate planet (*11*).

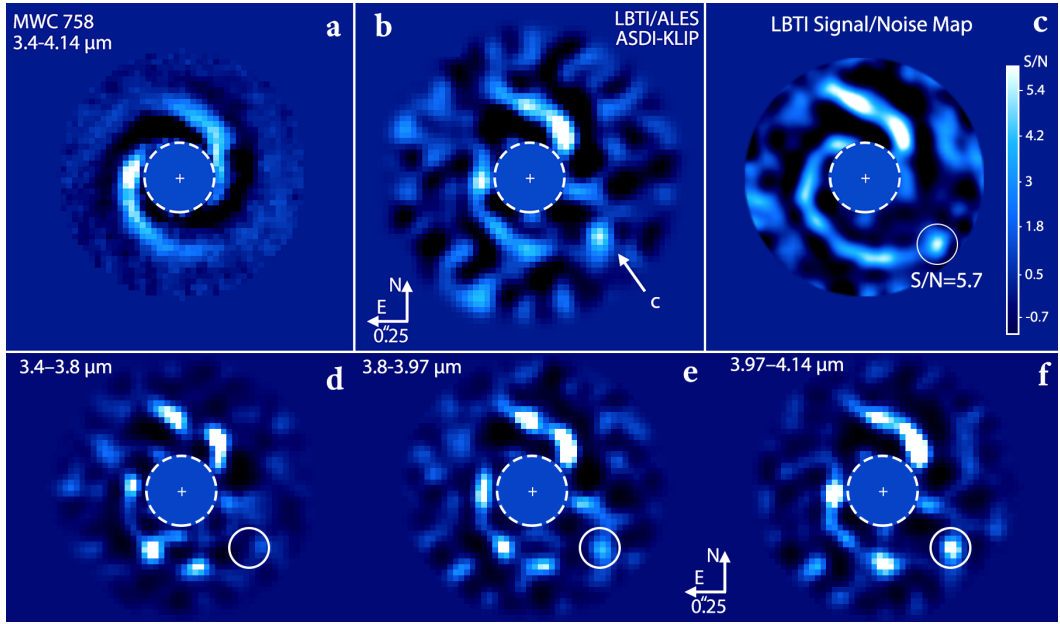

**Figure 1**. Images of MWC 758 from LBTI/ALES taken on UT 2019-11-14. **a:** data processing optimized for recovering extended structures (see Methods). **b:** the same wavelength range with processing optimized for recovering point sources. Note that injected planets with the brightness of MWC 758c are not recovered in the processing for extended sources–i.e., the non-detection of MWC 758c in panel **a** is not surprising. **c:** combined signal to noise map from all LBTI data (including $L'$ and $M'$ LMIRCam data published in *11*). **d-f:** data processing for point source recovery in narrower spectral bins, showing relatively constant flux from the disk and a notable rise of flux from the planet candidate toward longer wavelengths. Note that these images are processed with angular differential imaging, and thus the apparent features within the disk (especially in the reductions optimized for point sources: panels **b-f**) should not be interpreted as accurate representations of the disk's surface brightness.

MWC 758 (d=156 pc, age=3.5±2 Myr, SpT=A8Ve; *12, 13*) is among a subset of circumstellar disks that shows a two-armed spiral pattern but no obvious signs of a stellar or brown dwarf companion (*11, 14-16*; see also Fig. 1). The spirals display a clear two-armed geometry with an arm-to-arm separation of 140–170° (*11*), which can be linked to theoretical models of companion-driven spiral structures to imply a mass ratio of $q \gtrsim 0.005$ relative to the central star (i.e., $M_{companion} \gtrsim 8\ M_{Jup}$; *17*). The pitch angle of the spiral arms (~25–29°), which is linked to disk mass (*19*), suggests that the disk-to-star mass ratio is low enough that the disk is stable to its own self-gravity (*11, 20*). Likewise, the pattern speed measurements (*21, 22*) are incompatible with those of spiral arms formed by self-gravity. This supports the companion-driven hypothesis. The reported spiral arm rotation rate of 0.22°/yr±0.06°/yr (when including inclination: *20*) corresponds to a companion at ~$160^{+35}_{-25}$ au for a distance of 156 pc and a stellar mass of 1.5 $M_\odot$. The uncertainty was derived from measurements that oversampled the spatial resolution without accounting for the measurement covariance. Consequently, the uncertainty is underestimated.

Previous direct imaging observations of MWC 758 have revealed two candidate planets: MWC 758b, interior to the spiral arms at 0.11" or ~17 au projected separation (*16*), and MWC 758 CC1, for 'Companion Candidate 1', which is exterior to the Southern arm at 0.62" or ~97 au (*11*). Either candidate planet, if real, would be massive enough to generate the spiral arms. MWC 758b was not recovered in subsequent observations (*11*), whereas the prior observations (*16*) were not sensitive enough to detect MWC 758 CC1, leaving its nature unclear. Based on the spatial density of objects with similar infrared brightness, the possibility that MWC 758

CC1 could be an unassociated background object is <1% (see Methods). However, (*11*) could not exclude the possibility that CC1 could be a spurious detection, as the first few available detections established a combined ~10% false positive probability based on the residual speckle density distributions. In this work, we aim to resolve the nature of this candidate.

## Results

We observed MWC 758 with the Large Binocular Telescope Interferometer (LBTI; *23*) on UT 2019-01-05 and 2019-11-14. Observations were performed with the LMIRCam camera (*24, 25*) utilizing the Arizona Lenslets for Exoplanet Spectroscopy (ALES; *26, 27*). With 0.6 hr of integration time (source+sky, with equal time each, and seeing of ~1") on 2019-01-05, we first measured a very red, but low-SNR, spectrum of CC1 between $\lambda$=3.4–4.2 µm that appeared to be distinct from the rest of the disk. To verify this result, we obtained a longer observation of 1.8 hr integration on 2019-11-14 in photometric conditions (~0.8–1.0" seeing). The results verified our initial findings and improved the SNR from an average value of ~1 in each spectral channel to ~3 (for $\lambda \geq 3.8$ µm). The spectral resolution of both observations was $R$~40, enabling features as small as ~0.1 µm to be identified.

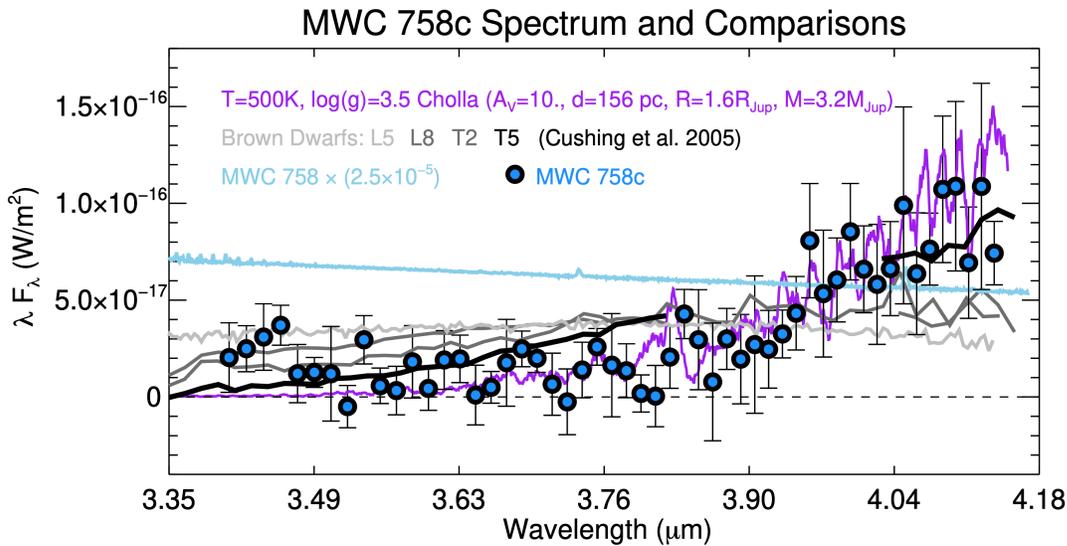

**Figure 2.** LBTI/ALES spectrum of MWC 758c (blue points) combined with a simple average over the two epochs. Comparisons of L5–T5 brown dwarf spectral standards are shown in gray through black (*29*), and an example of a cold and moderately attenuated Cholla model atmosphere is shown in purple (*30*). Note that higher (lower) levels of assumed extinction correspond to higher (lower) inferred effective temperatures. The IRTF/SpeX spectrum of the star and warm inner disk is shown in the light blue curve. The scattered light spectrum from the spiral arms closely resembles this source spectrum (see Methods for further comparisons).

To enable conversion of contrast measurements to physical flux units, we also obtained a flux-calibrated spectrum of MWC 758 with NASA's Infrared Telescope Facility (IRTF) / SpeX instrument (*28*) on 2021-02-03 (see Methods and Fig. 2). In summary of the findings from LBTI/ALES and IRTF/SpeX, MWC 758 CC1 has a spectrum that is consistent with a very faint and very red point source, whereas the rest of the disk has a spectrum consistent with scattered starlight (see Fig. 2 and Methods)–this is even more apparent with observations of the disk at shorter wavelengths taken into account (e.g., *15*). Therefore, we determine MWC 758 CC1 to be an object distinct from the rest of the disk and refer to this object henceforth as MWC 758c.

To expand the spectral range, we analyzed archival data covering $\lambda = 0.95$–2.2 µm from the Very Large Telescope (VLT)/Spectro-Polarimetric High-contrast Exoplanet REsearch (SPHERE: *31*) instrument. This enabled us to place constraints on the level of extinction (or $V$-band attenuation, $A_V$) when compared to the detections from LMIRCam and ALES at $\lambda = 3$–5 µm. The SPHERE data yield a non-detection of MWC 758c (upper limit ~5–6 × $10^{-6}$ contrast at $J$– and $H$-band and SNR~3, see Methods), while the disk is detected with a similar brightness and morphology to its appearance in the images at $\lambda \sim 3$–5 μm (*16*, Fig. 1). This near-IR non-detection of MWC 758c places a lower limit of $A_V \geq 8$ for MWC 758c (see Methods). The star has an extinction of $A_V = 0.4$, which suggests that interstellar extinction is negligible (*14*), and that the attenuation is likely originating from circumplanetary dust, since both the μm-sized scattered-light (*15*) and emission from mm-sized circumstellar dust disk (*32*) fall off sharply near the projected separation of MWC 758c.

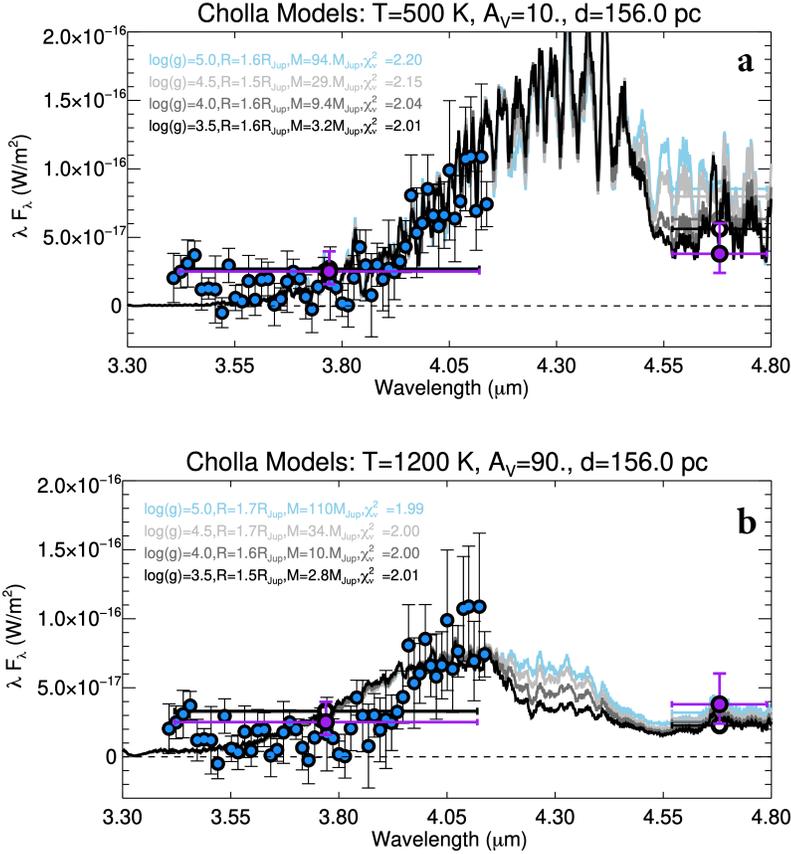

**Figure 3.** Comparison of LBTI/ALES data (blue points) and broadband photometry from LBTI/LMIRCam (purple; *11*) to Cholla model atmospheres (light blue, gray, and black curves, *30*). Synthetic LMIRCam photometry are shown in light blue, gray, and black points. Panel **a** shows an example for a case of low $A_V$ and low $T_{\text{eff}}$, whereas panel **b** shows an example for a case of high $A_V$ that requires a higher $T_{\text{eff}}$, corresponding to typically more massive planets. JWST's NIRISS instrument will be able to be able to fill in the range between 4-5 μm, which will constrain $A_V$ (and $T_{\text{eff}}$) through the strength of the molecular absorption features.

We compared the spectral measurements of MWC 758c to standard brown dwarfs and planetary atmosphere models (*29*, *30*, *34*) in order to estimate its physical properties. The brown dwarf comparison yields a best match of a very late spectral type of T5 or later, consistent with a relatively cold object (see Fig. 2). The atmospheric model comparison yields constraints on the combination of effective temperature ($T_{\text{eff}}$), radius, and level of attenuation ($A_V$). Assuming radii between $R = 1 - 2$ $R_{Jup}$ (motivated by the predicted range from evolutionary models for the system's age of ~3.5 Myr: e.g., *34*, *35*), a reddening relation typical for interstellar dust (*36*), and a range of $A_V$≤150 (motivated by simulations of planets forming embedded within protoplanetary disks: *37–39*), the brightness of MWC 758c is consistent with temperatures between $T_{\text{eff}} = 400 - 2500$ K (Figs. 3, 4). Higher values of $A_V$ correspond to much higher values of $T_{\text{eff}}$ for a given radius (Fig. 4). The starting temperatures of low initial entropy, or "cold-start" models is uncertain, but typically considered to be $T_{\text{eff}} = 600 - 800$ K (*35*, *40*). The surface gravity ($g$) and mass are relatively unconstrained, although the best-fitting solutions favor lower values (see Methods). Finally, the most likely inferred ranges of temperature and radius are consistent with those of the expected body driving the spiral arms ($M \gtrsim 8$ $M_{Jup}$: *17*) from both cold– and hot-start models of mass, temperature, and radius evolution (*34*, *35*, *41*, *42*).

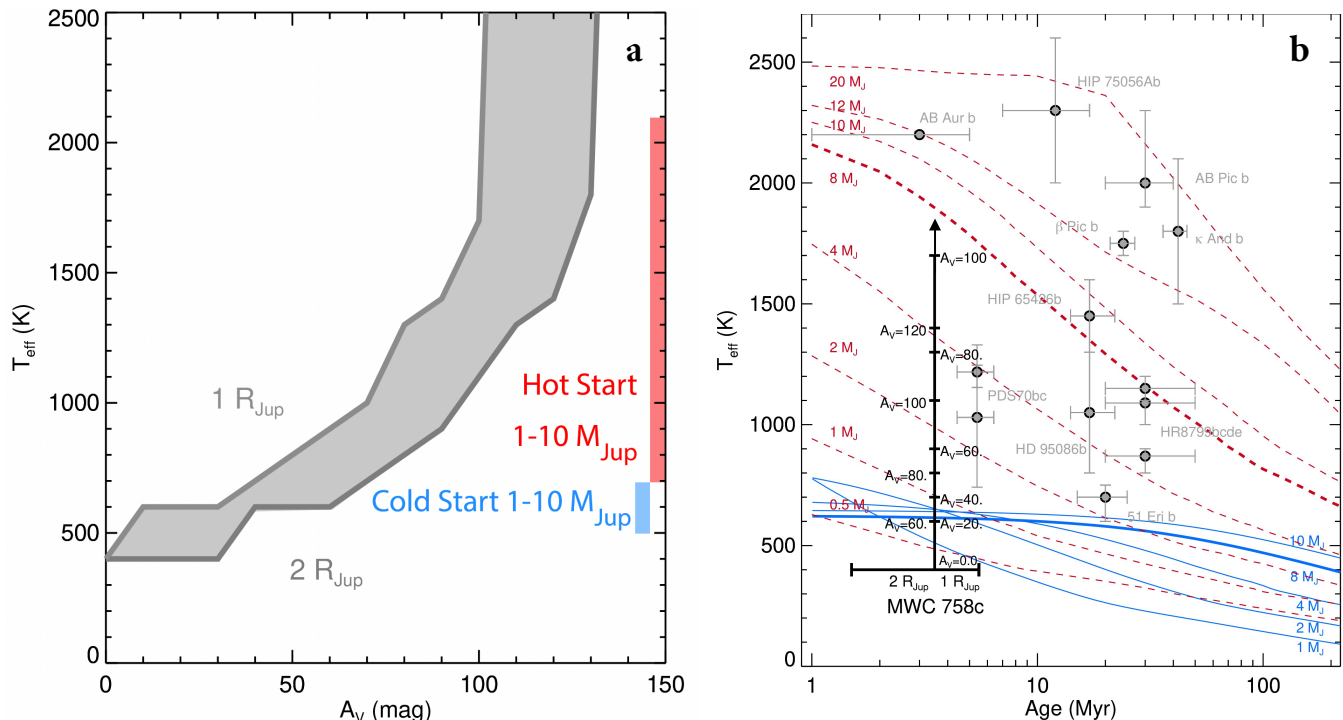

**Figure 4. a** Retrieved $T_{\text{eff}}$ vs. $A_V$ of MWC 758c for a range of plausible planetary radii. Hot-start models (e.g., *34*) show planets with maximum radii up to ~2 $R_{\text{Jup}}$, whereas cold-start models (e.g., *35*) have radii closer to ~1 $R_{\text{Jup}}$. Either set of models is consistent with the data for a reasonable range of visual extinction and planet mass. **b** Effective temperature vs. age of directly imaged exoplanets and brown dwarf companions compared to evolutionary tracks for hot– (red: *42*) and cold-start planets (blue: *35*). MWC 758c's temperature and mass are not well constrained due to the degeneracy with $A_V$. Nevertheless, MWC 758c stands out both as one of the youngest giant planets to be directly imaged (indeed, among just a few that are known within the protoplanetary disk phase), and also as either one of the coldest or most attenuated.

## Discussion

### Evidence For MWC 758c Being a Planet

With this work, we have established that MWC 758c has a spectrum that is distinct from the rest of the disk (see Methods) and consistent with that of a planetary atmosphere. MWC 758c also does not appear in polarized emission (*15*), while the spiral arms bear the same morphology in total intensity and polarized light (*14–16*). This suggests a self-luminous object rather than scattered starlight. Combined with the non-detection of MWC 758c at wavelengths shorter than $\lambda \leq 3.3$ μm (compared to obvious detections of the disk: *15*, *18*), the data are only consistent with a reddened thermal emission spectrum.

Our reddening constraint for MWC 758c ($A_V \geq 8$) is consistent with predictions from hydrodynamical simulations ($A_V \sim 10 - 150$: *38*, *39*) and also with the estimates for the PDS 70 protoplanets ($A_V \sim 16 - 17$: *43*, although *44-46* find smaller values of $A_V \sim 1 - 8$). AB Aur b is also likely in an embedded phase (*47*). The few known examples of imaged protoplanets, including MWC 758c, are all consistent with having significant levels of optical attenuation, as predicted by models of giant planet formation (e.g., *37–39*). We also note an additional possibility: that the emission observed from MWC 758c could be either partly or entirely due to a circumplanetary disk or envelope, with the planet itself completely obscured or too faint to be detected. Compared to models in which the circumplanetary disk outshines the planet itself (*37*), the ~5 $M_{Jup}$ model provides a reasonable fit to the rise in brightness at $\lambda \gtrsim 3.8$ μm. Due to the probable low mass of MWC 758c (comparable to a thermal mass; see below), the circumplanetary material may also more closely resemble an envelope than a disk (*47*), leading to greater expected levels of attenuation of the protoplanet's own emission.

Dynamical effects that are expected from a forming giant planet are seen in the disk around MWC 758. In particular, the large-scale spiral pattern cannot be attributed to gravitational instability nor a recent stellar fly-by (e.g., *49*, *50*) and is likely companion driven. Comparing the measured separation of the spiral arms to predictions from theoretical models, the companion is expected to have a mass ratio with respect to the central star of q>~0.005 (*17*), or $\gtrsim 8$ $M_{Jup}$ assuming a central star of 1.5–1.9 $M_\odot$ (*51*). This is consistent with the spectroscopically inferred mass of MWC 758 with $A_V \sim 100$ for hot-start planets, or $A_V \gtrsim 10$ for cold-start planets.

The spectroscopically constrained mass of MWC 758c likely straddles (or slightly exceeds) the thermal mass at $a$~100 au: $M_{thermal} = M_\star(h/r)^3$, where h/r is the disk aspect ratio, or a few Jupiter masses assuming h/r ~ 0.1. Therefore, the planet is expected to remove gas and dust from its co-orbital region and could possibly open a gap (*33*). The extent of this process and the resulting gas and dust depletion depend on poorly constrained parameters such as the disk viscosity and orbital parameters of the planet including eccentricity and inclination (*33*). Whereas the excitation of the spiral arms occurs on the dynamical timescale (~$10^3$–$10^4$ yr: *52*), the depletion of material around the planet's orbit occurs on the viscous timescale, which can be comparable to the system's age even for a modest viscosity ($\alpha \gtrsim 0.001$). This further limits our ability to quantitatively predict the outcome of gap opening for MWC 758c, leaving open the possibility for gas and dust to be present within the planet's co-orbital region.

Nevertheless, ALMA observations of the gas and mm-sized dust provide evidence consistent with the disk responding to the planet's gravitational perturbation in addition to the spirals. The mm-dust disk falls off sharply near the projected separation of MWC 758c, while the $^{13}CO$ gas and $C^{18}O$ gas are also in steep decline (*32*, *53*). The latter also hints at the likely presence of an inner gap edge, which is needed to explain the observed emission clump at 0.53 arcsec to the North of the star, assuming that the clump is a vortex triggered by the Rossby wave instability (*54*, *55*). The proximity of the planet to the outer edge of the mm-dust and $^{13}CO$ gas is not predicted by current steady state models of disk-planet interactions if MWC 758c is above the thermal mass (e.g., *33*). Future constraints on the formation timescale of the planet, and also on planet mass (which will soon be possible through constraints on $A_V$ from JWST), will enable an assessment of this possible discrepancy with current models. As an aside, we note that MWC 758c, being exterior to the inner cavity, is likely not responsible for the structures interior to ~0.3 arcsec.

**Implications of MWC 758c for Other Systems**

As the reddest known planet and one of few known protoplanets, the existence of MWC 758c has a number of implications for the broader population of spiral protoplanetary disks and young giant planets. Perhaps most notably, the high level of optical extinction ($A_V \geq 8$) indicates the likely presence of a circumplanetary disk around MWC 758c, suggesting also that accretion is likely ongoing, despite the non-detection of H$\alpha$ (*9*, *10*). As MWC 758c is among the most attenuated planets yet to be detected (perhaps the most attenuated), it provides an opportunity to constrain the properties (e.g., grain size and chemical makeup) of the line-of-sight material. Existing reddening laws are derived primarily from interstellar extinction and star forming galaxies, whereas grain growth and processing within circumstellar and circumplanetary disks may result in differences in the chemical makeup of the material accreted during planet formation and a different reddening profile.

Second, MWC 758c confirms that spiral arms in protoplanetary disks can be driven by giant planets that are cold enough ($T_{eff} \lesssim 600$ K) or reddened enough ($A_V \gtrsim 10$) to have escaped detection at shorter wavelengths ($\lambda \sim 1 - 2$ μm). Indeed, this was predicted (*7*) based on the status at the time that no giant planets had been found around disks with prominent spiral arms. The finding of MWC 758c serves as a proof-of-concept that spiral arms in other protoplanetary disks (e.g., SAO 206462: *3*) may also be caused by a class of faint and very red planets that are more readily detectable at mid-infrared wavelengths by systems like

LMIRCam/ALES. Relatively few surveys have been performed at mid-IR wavelengths compared to those in the optical to near-IR, which would be blind to planets as red as MWC 758c. Now, the James Webb Space Telescope (JWST) is also capable of imaging fainter and even more attenuated planets around a greater number of stars. MWC 758c is resolvable by JWST for $\lambda \lesssim 10$ μm. Such data could place better constraints on $T_{\text{eff}}$ and $A_V$ via the strength of molecular absorption features, and can probe accretion-tracing Hydrogen emission at $\lambda = 4.05$ μm (Brackett-$\alpha$). With its stable observing environment, JWST may also detect brightness variations in MWC 758c caused by variable attenuation from dust orbiting (or accreting onto) the planet on dynamical timescales of a day or less, which could help to better constrain the reddening law specific to the dust around MWC 758c.

Finally, for low levels of attenuation ($A_V \lesssim 10$) MWC 758c would be the coldest currently known directly imaged planet (the closest planet in temperature would be 51 Eri b with $T_{\text{eff}} = 600 - 750\ K$: *56*). At this colder end of the plausible range of temperatures for MWC 758c, methane is the dominant carrier of atmospheric carbon (e.g., *29, 57*), and at the coldest temperatures considered ($T_{\text{eff}} \sim 400K$) water clouds are possible (*57, 58*). A planet with $T_{\text{eff}} \sim 400K$ would occupy an intermediate temperature range between Jupiter and the predominantly warmer and CO-dominated directly imaged exoplanets. If at the colder end of its range of $T_{\text{eff}}$, MWC 758c could provide one of the first opportunities to study the carbon chemistry of these colder exoplanet atmospheres, which so far have only been studied via isolated brown dwarfs (e.g., *57, 58*).

## Conclusion

We have presented images and spectroscopy of a probable young giant planet driving the spiral arms in the circumstellar disk of MWC 758. The spectrum of MWC 758c is consistent with that of a very red protoplanet–either due to a low effective temperature ($T_{\text{eff}} \leq 600$ K, which given the system's young age of 3.5±2 Myr would imply a cold-start origin: e.g., *35*) or significant attenuation by dust ($A_V \geq 8$). Furthermore, the spectrum of MWC 758c is distinct from that of the rest of the disk and is inconsistent with scattered starlight. The existence of MWC 758c has two important implications: 1) faint and red giant planets are capable of driving large-scale spiral structures in protoplanetary disks, and 2) protoplanets in a similar phase of evolution to MWC 758c (in particular, with similarly red spectra) would likely have been missed by past surveys–including those performed in the accretion tracing $H\alpha$ filter (*59*). Mid-IR observations, like those presented here, and those that are now possible with JWST, are able to reveal these young (and likely very red) protoplanets.

Correspondence and requests for materials should be addressed to K. Wagner (kevinwagner@arizona.edu).

## Methods
### LBTI/ALES Observations and Data Reduction
We processed the data for each night in a nearly identical manner following the reduction strategy for LMIRCam data in (*11*). We briefly recount this process here–mostly focusing on the differences in the approach for Integral Field Spectroscopy (IFS) data. Note that the early calibration steps are very similar, since ALES utilizes LMIRCam's full optomechanical chain and detector (*24, 25*) with the addition of a lenslet array in the optical path (for more details, see *26* and *27*). The data reduction procedure is described below. Parameters are given for the UT 2019-11-14 epoch and for the (UT 2019-01-05) epoch in parentheses where they are different.

1. We removed the thermal and instrumental background by subtracting the average of the neighboring nod positions from each frame. Each observation used 300 frames per nod.

2. We removed reset noise by subtracting the first read of each ramp from the final read (correlated double sampling). The total exposure time of each single LMIRCam exposure was 1.967 sec (0.984 sec). We obtained a total of 3230 (2196) frames, including sky exposures for background subtraction that accounted for 50% of the total observing time. This amounts to 0.88 (0.30) hr of on-source exposure time, excluding overheads.

3. We subtracted the bias of each vertical channel by measuring and subtracting the average value of the pixels in the overscan region.

4. We extracted 3-dimensional (x-y-$\lambda$) data cubes from each individual frame (following *60–62*).

5. We identified wavelengths via sky frames taken through four narrowband ($R$~100; 2.897µm, 3.36µm, 3.539µm, and 3.874µm) filters (see Section 4 of *27*).

6. We selected bad pixels as those that are 2-$\sigma$ outliers and/or those with a value less than –100 in a 5 pixel × 5 pixel square (ignoring the target pixel) in a median image of the first 50 frames at each wavelength. We tested thresholds of 3-$\sigma$ and 4-$\sigma$ and found consistent overall results. These resulted in ~2%, 1%, and 0.5% of pixels being flagged as bad, respectively.

7. We replaced the bad pixels via interpolation of the surrounding pixels (up to a maximum target distance of 3 pixels from the target pixel).

8. We centered the frames by first fitting the maximum of the peak, and subsequently by computing the position of maximum correlation of the frames with the first in the sequence.

9. We rejected bad frames as those with a maximum correlation of less than 0.99 (0.96) with respect to the median image (averaged over wavelength). This resulted in 0.3% (18%) rejection of frames. The higher rejection fraction on the 2019-01-05 epoch reflects the poorer seeing conditions.

10. When relevant, we injected synthetic point sources at this stage. For a PSF template, we used the median image of the star at each wavelength, which remained unsaturated and in the detector's linear regime.

11. We generated derotated, derotated+high pass filtered, high pass filtered + classical angular differential imaging (ADI: *63*), and high pass filtered + ADI-KLIP processed images (KLIP stands for Karhunen-Loève Imaging Processing: *64*). Throughout the observation, the parallactic angle evolved by 127º (119º). For the 2019-01-05 epoch, at this stage we first destriped the images by subtracting the mode from each column and row of pixels. We used a high pass filter width of 11 (9) pixels and tested a wide range of KLIP hyper parameters (namely, the distribution of subregions and numbers of KLIP components, or KL modes). Before KLIP processing, we binned the data by averaging over each set of 15 (10) frames in the sequence. We generated two KLIP images: one with a less aggressive set of parameters that leaves the disk structures largely intact, but is less sensitive to point sources; and one with a more aggressive set of parameters with greater sensitivity to point sources. For the less aggressive parameters, we used full annuli between 5–27 pixels (0.175–0.945 arcsec at 35 mas/pixel) and five KLIP components, with additional parameters identical to those as follows. For the more aggressive set of parameters, we split the annulus into 6 (8) annular segments of equivalent width, and used 15 KLIP components. Injected point-sources with the brightness of MWC 758c are not detected in the less aggressive reduction, but are in the more aggressive one (see Fig. S1). For constructing the basis of KLIP eigenimages we rejected frames taken with a parallactic angle within 0.1 degree (0.85 degree) of the target image in order to reduce self-subtraction of point sources.

12. To further reduce wavelength-dependent speckles in the final images, we generated spectral differential imaging (SDI)-KLIP processed images from the ADI-KLIP processed cubes (i.e., ASDI-KLIP processed images in the end). For SDI-KLIP parameters, we used the same regions as above, with a rejection criteria of frames whose wavelength ratio (i.e., magnification ratio) resulted in less than 1.5×FWHM of separation for point sources at the center of the radial processing range.

13. We high-pass filtered the images a second time, and then combined images at each wavelength with a variance-weighted combination (*65*).

**VLT/SPHERE Observations and Data Reduction**
Data were taken on two separate nights and in two separate modes: with the integral field spectrograph (IFS) arm operating between *Y*– to *H*-band ($\lambda \sim 0.95$–1.65 μm) and the dual-band imager (IRDIS) operating in the *K12*-bands ($\lambda \sim 2.11$ and 2.25 μm) on UT 2016-01-01 under program-ID 096.C-0241 (PI: Beuzit), and with the IFS operating between *Y*– to *J*-band ($\lambda \sim 0.95$–1.35 μm) and IRDIS operating in the *H23*-bands ($\lambda \sim 1.59$ μm and 1.67 μm) on UT 2018-12-17 under program-ID 1100.C-0481 (PI: Beuzit). The data quality on the second night was overall substantially higher. For data processing details, we followed the methodology of *66*, which is very similar to that described above for LBTI/ALES. We refer to this prior work and the above description for further details. The images are shown in Fig. S2 and Fig. S3. At the separation of MWC 758c, the *YJ* image provides the deepest detection limits of $\sim 3 \times 10^{-6}$ contrast (with SNR~3 and for a flat spectrum over the complete bandpass; see panel d of Fig. S3). Detection limits in narrower synthetic photometric bands are somewhat higher (e.g., $\sim 6 \times 10^{-6}$ contrast in *J*-band with SNR~3). The IRDIS-H23 dataset (taken simultaneously) provides a similar limit of $\sim 5 \times 10^{-6}$ contrast with SNR~3. The ghost in the YJ images is due to persistence in the detector following flux calibration, and is identifiable via its rapidly diminishing brightness with frame progression and regular appearance in other datasets.

**IRTF/SpeX Observations and Data Reduction**
We observed MWC 758 with IRTF/SpeX (*28*) on UT 2021-02-03 using the cross-dispersed (XD) echelon gratings in both short (SXD) and long (LXD) wavelength modes, covering 0.8–2.4 μm and 2.3–5.4 μm, respectively, and reduced the data using Spextool (*67*). A 0.8 arcsec slit was used during the XD observations, which amounted to a small amount of light loss due to average seeing of 0.5 arcsec. The LXD and SXD spectra were normalized in the region of overlapping wavelength coverage (2.3–2.4 μm). Observations of an A0V spectral standard (HD 34203) were used for telluric correction and flux calibration, and separate observations of MWC 758 through a low-resolution prism with a 3 arcsec slit were used to measure a ~14% absolute correction to the XD measurements to account for the light lost at the smaller (0.8 arcsec) slit. The spectra are shown in the next subsection at the stage at which they are used to correct the LBTI/ALES contrasts (step 5)

**LBTI/ALES Spectroscopy**
We extracted and analyzed the spectra from each night following a uniform approach:

1. We began by extracting the spectrum of MWC 758c, of ten different locations along the spine of the spiral arms (beginning at their furthest visible extent), and of the star, all within apertures of 3 pixels in diameter. We checked that using various aperture diameters between 1–4 pixels provides consistent results. These are presented in Fig. S4 and Fig. S5.

2. We compared the spectra of MWC 758c and the disk for each night independently. The averaged disk spectra are presented in Fig. S5.

3. We cross-correlated each pixel's spectrum vs. a $T_{\text{eff}} = 500$K, $A_V = 40$ BT-Settl spectrum (*34*) for the higher-quality 20191114 dataset to create a correlation map (Fig. S6). The measured spectra were smoothed by a running median of 5 spectral channels (corresponding to the spectral resolution, *R*~40). The image then was spatially smoothed by a 2 pixel boxcar filter, which is approximately the FWHM. Prior to smoothing, MWC 758c has a maximum cross correlation of 0.95 with respect to the template spectrum. The rest of the image on average, including the spiral disk (shown in contours in Fig. S6 along with MWC 758c), has a maximum cross correlation of 0.55 and a standard deviation of 0.15–i.e., MWC758c's red spectrum is a ~3 standard deviation outlier from the spectra contained within the rest of the image, and also compared to just those pixels containing scattered light from the circumstellar disk. We note also that we did not use the best-fit spectrum, but simply a relatively cold/red atmosphere as a comparison. We verified that any $T_{Eff}$=400–1000K atmosphere provides similar results. The rest of the spiral arms, if fit to thermal emission spectra, return much higher values (up to the maximum of the tested range of 2000K). This is unsurprising since the spectra are the result of scattered starlight.

4. We performed forward modeling via synthetic point source injections (e.g., *68*) to generate a point source flux correction model and to estimate uncertainties via the following procedure. We scaled the band-averaged contrast spectra to match the $L'$ contrast measured with LMIRCam (*11*) and injected point sources with this spectrum into the data at three separate locations (we also checked that using a flat spectrum does not significantly change the results). We similarly extracted their spectra and computed a linear correction model to the ratio of measured to injected fluxes (ignoring outliers as those with a correction factor greater than 10 or less than 0). The point source flux correction models are presented in Fig. S7. We also computed uncertainties as the standard deviation of the three measured values at each wavelength for the injected sources.

5. We created corrected contrast spectra of MWC 758c from the above correction models (Fig. S8). This improved the agreement between the two epochs and the previous $L'$ measurement. The simple average of the two epochs and their respective signal to noise ratios (SNR) are shown in Fig. S9. The SNR of the second epoch is higher than the first and of their combination (with errors propagated through combination). However, to avoid biases introduced by a single observation (particularly due to correlated speckles: see *69*), we proceed with the combined spectrum for the proceeding analysis. We verified that using either single night yields consistent results with those obtained using their average.

6. We converted the contrast spectra to units of W/m² by multiplying by the empirical spectrum of the star measured with IRTF/SpeX on UT 2021-02-03 (shown in Fig. S10). Note that this spectrum was taken at a different time than the LBTI/ALES data. Intrinsic variability of the central star and inner disk is thus a potential source of uncertainty (likely less than 20% based on the source's historical variability; *14*).

**Comparison to Atmospheric Spectral Models**

We compared the spectrum of MWC 758c to four different model grids: Cholla (*30*), BT-Settl (*34*), COND (*42*), and DUSTY (*70*). COND and DUSTY were among the first grids available for clear and cloudy atmospheres, respectively. These are outdated and included primarily for comparison to other previous works that used these grids. Additionally, the DUSTY grid is likely a poor match at cold temperatures since the clouds in the models don't settle out. For cloudy (cloudless) planets, the BT-Settl (Cholla) models are the most reliable.

For a given $A_V$, and with a reddening relationship of $A_V = 3.2E(B − V)$ from *36*, the script first scales the model spectra to the ALES data (normalized to the mean flux of the bandpass) and then calculates a radius based on that normalization and the Gaia DR3 distance to the system. When maximum radius drops below the set threshold (1 or 2 $R_{Jup}$ for simplicity in Fig. 4, but ~1.1 and 1.7 $R_{Jup}$ are closer to the range of models) for a given temperature and range of $log(g)$, the script determines that temperature to be the maximum plausible temperature for the prescribed $A_V$. The age is not included in this analysis since the goal is to find the plausible range of temperatures that can then be used to compare to the $T_{eff}$, mass, radius, and age framework in the right panel of Fig. 4. To assess the model-data fit, we computed the chi-squared ($\chi^2$) metric for the 54 spectrally-correlated ALES datapoints, accounting for the spectral covariance (*69*). The $\chi^2$ value from the ALES data was added to the two independent $\chi^2$ values from the $L'$ and $M'$ measurements and converted to reduced chi-squared by dividing by $(n − 3) = 53$, accounting for the three free parameters of $T_{eff}$, $A_V$, and $log(g)$. The results of this analysis are shown in Fig. S11. Note that the *M*-band photometry is driving the gravity fit for the BT-Settl, AMES-Cond, and AMES-Dusty models; however, for brown dwarfs in this temperature range, non-equilibrium chemistry plays a larger role than surface gravity on *M*-band photometry (*71*)–therefore, the (already weak) constraints on $log(g)$ are even less constrained when accounting for a range of vertical mixing. The Cholla models (that do include vertical mixing) shown are for $log(K_{zz})$=4 (the middle of the available range). The other values, $log(K_{zz})$=2 and $log(K_{zz})$=7, provide similar results, with slightly higher minimum $\chi^2_\nu$ values for $log(K_{zz})$=2 and slightly lower for $log(K_{zz})$=7 (i.e., a greater degree of vertical mixing is favored for the cloud-free Cholla models). Overall, the BT-Settl models produce the best match.

**Possibility that MWC 758c is a background object**
We considered the hypothesis that MWC 758c could be a background object that is projected behind MWC 758 but otherwise unassociated with the system. We note *a priori* that this is unlikely due to the dynamical requirement of a companion responsible for the spiral arms in MWC 758, which are easily explained if MWC 758c is indeed a gravitationally bound component of the system. Beyond this initial line of reasoning, we consider 1) common proper motion, which is not conclusive due to the level of astrometric uncertainty, and 2) the spatial density of objects with a similar *L′* brightness, which suggests a <1% chance of such an occurrence.

MWC 758 has a proper motion of –26 mas/yr in declination and 4 mas/yr in right ascension (from *23*, rounded to the nearest mas, and with negligible uncertainties). Over the three year baseline of our observations (taking the first epoch of UT 2016-10-15 from *11*), we would expect to see a relative motion of MWC 758c with respect to MWC 758 of ~80 mas, assuming a zero proper motion background object. The relative position of MWC 758c in the former epoch was 0.617±0.024 mas in separation and with a position angle of 224.9±2.2°E of N. In our latest ALES epoch (UT 2019-11-14), we measure a separation of 0.603±0.0345 arcsec and 228.5±3.2° E of N. We conservatively consider the uncertainty of the ALES astrometry as ±1 pixel, or 34.5 mas, since a typical estimate for the uncertainty of FWHM/SNR yields ~30 mas. We measure a relative motion of 41±59 mas between 2016–2019. This is consistent with zero observed motion and closer to the expected motion of a giant planet on a circular orbit at 100 au (~20 mas), but is not precise enough to confidently rule out a stationary background track. However, the images here have not been calibrated against an astrometric field, and given that the lenslet array is a non-stationary optic, it would not be surprising for the systematic astrometric uncertainty to be larger than the uncertainty quoted here. Therefore, we cannot confidently claim that we have measured an actual motion of MWC 758c.

Next, we consider the possibility that MWC 758c is a background object based on the spatial density of objects with similar *L′* brightness. In a recent survey with LBTI/LMIRCam (i.e., using the same system with the same background-limited sensitivity), just four background objects were found among 98 targets (*72*), each with a circular field of view of 3" in radius, and including the Taurus star forming region (of which MWC 758 is a likely member). This translates to a spatial density of 0.0014 sources/arcsec$^2$. Due to the effects of tidal truncation of a very massive companion on the disk, we would have only identified a source as plausibly responsible for the spiral arms if it were within a few times the separation of the spirals, or ~1.5". Thus, over a 1.5" circular field of view, the false alarm probability of MWC 758c is ~1%.

## Data Availability
All LBTI data are in the process of being integrated into the LBTO archive (http://archive.lbto.org). The data for this program are not yet available online, but in the future will be accessible by searching for PI: Wagner or Target: MWC 758. Until then, the raw and processed data will be available upon request of the authors.

## Code Availability
All software, codes, and data processing scripts that were developed for this study are available at https://github.com/astrowagner/MWC758_ALES.

## Acknowledgements

We thank Theodora Karalidi, Judit Szulágyi, Thayne Currie, Rachel Fernandes, Bin Ren, and Chengyan Xie for conversations that were helpful to our analysis. Support for this work was provided by NASA through the NASA Hubble Fellowship grant HST-HF2-51472.001-A awarded by the Space Telescope Science Institute, which is operated by the Association of Universities for Research in Astronomy, Incorporated, under NASA contract NAS5-26555. This paper is based on work funded by NSF Grant no. 1608834. The results reported herein benefited from collaborations and/or in- formation exchange within NASA's Nexus for Exoplanet System Science (NExSS) research coordination network sponsored by NASA's Science Mission Directorate. MLS, JW, and KL were supported in part by NASA XRP program via grants 80NSSC20K0252 and NNX17AF88G. FM acknowledges funding from the European Research Council (ERC) under the European Union's Horizon 2020 research and innovation program (grant agreement No. 101053020, project Dust2Planets). We acknowledge the expertise of the LBTO staff, including Jennifer Power and Jared Carlson, for their support of these observations, and to LBTO director Christian Veillet for enabling these observations through director's discretionary time.

## Author Contributions

Observations (KW, JS, SE, ES, MS, AB, KA, AB, RR); Data Analysis (KW, JS, AS); Interpretation of Results and Preparation of Manuscript (KW, JS, AS, SE, RD, DA, ES, JL, MS, KK, TB, MM, BM, AB, KA, AB, TU, CW, PH, ZB, KL, FM, EP, RR, MS, JW).

## References

1. Huang, J., Andrews, S. M., Dullemond, C. P., Isella, A., Pérez, L. M., Guzmán, V. V., Öberg, K. I., Zhu, Z., Zhang, S., Bai, X.-N., Benisty, M., Birnstiel, T., Carpenter, J. M., Hughes, A. M., Ricci, L., Weaver, E., & Wilner, D. J. (2018). *The Disk Substructures at High Angular Resolution Project (DSHARP). II. Characteristics of Annular Substructures.* ApJL, 869, L42.
2. Benisty, M., Dominik, C., Follette, K., Garufi, A., Ginski, C., Hashimoto, J., Keppler, M., Kley, W., & Monnier, J. (2022), *Optical and Near-infrared View of Planet-forming Disks and Protoplanets.* arXiv e-prints, arXiv:2203.09991.
3. Muto, T., Grady, C. A., Hashimoto, J., Fukagawa, M., Hornbeck, J. B., Sitko, M., Russell, R., Werren, C., Curé, M., Currie, T., Ohashi, N., Okamoto, Y., Momose, M., Honda, M., Inutsuka, S., Takeuchi, T., Dong, R., Abe, L., Brandner, W., Brandt, T., Carson, J., Egner, S., Feldt, M., Fukue, T., Goto, M., Guyon, O., Hayano, Y., Hayashi, M., Hayashi, S., Henning, T., Hodapp, K. W., Ishii, M., Iye, M., Janson, M., Kandori, R., Knapp, G. R., Kudo, T., Kusakabe, N., Kuzuhara, M., Matsuo, T., Mayama, S., McElwain, M. W., Miyama, S., Morino, J.-I., Moro-Martin, A., Nishimura, T., Pyo, T.-S., Serabyn, E., Suto, H., Suzuki, R., Takami, M., Takato, N., Terada, H., Thalmann, C., Tomono, D., Turner, E. L., Watanabe, M., Wisniewski, J. P., Yamada, T., Takami, H., Usuda, T., & Tamura, M. (2012). *Discovery of Small-scale Spiral Structures in the Disk of SAO 206462 (HD 135344B): Implications for the Physical State of the Disk from Spiral Density Wave Theory.* ApJL, 748, L22.
4. Dong, R., Zhu, Z., Rafikov, R. R., & Stone, J. M. (2015). *Observational Signatures of Planets in Protoplanetary Disks: Spiral Arms Observed in Scattered Light Imaging Can be Induced by Planets.* ApJL, 809, L5.
5. Dong, R., Zhu, Z., Fung, J., Rafikov, R., Chiang, E., & Wagner, K. (2016). *An M Dwarf Companion and Its Induced Spiral Arms in the HD 100453 Protoplanetary Disk.* ApJL, 816, L12.
6. Wagner, K., Dong, R., Sheehan, P., Apai, D., Kasper, M., McClure, M., Morzinski, K. M., Close, L., Males, J., Hinz, P., Quanz, S. P., & Fung, J. (2018). *The Orbit of the Companion to HD 100453A: Binary-driven Spiral Arms in a Protoplanetary Disk.* ApJ, 854, 130.
7. Dong, R., Najita, J. R., & Brittain, S. (2018). *Spiral Arms in Disks: Planets or Gravitational Instability?* ApJ, 862, 103.
8. Bae, J., & Zhu, Z. (2018). *Planet-driven Spiral Arms in Protoplanetary Disks. II. Implications.* ApJ, 859, 119.

9. Cugno, G., Quanz, S. P., Hunziker, S., Stolker, T., Schmid, H. M., Avenhaus, H., Baudoz, P., Bohn, A. J., Bonnefoy, M., Buenzli, E., Chauvin, G., Cheetham, A., Desidera, S., Dominik, C., Feautrier, P., Feldt, M., Ginski, C., Girard, J. H., Gratton, R., Hagelberg, J., Hugot, E., Janson, M., Lagrange, A.-M., Langlois, M., Magnard, Y., Maire, A.-L., Menard, F., Meyer, M., Milli, J., Mordasini, C., Pinte, C., Pragt, J., Roelfsema, R., Rigal, F., Szulágyi, J., van Boekel, R., van der Plas, G., Vigan, A., Wahhaj, Z., & Zurlo, A. (2019). *A search for accreting young companions embedded in circumstellar disks. High-contrast Hα imaging with VLT/SPHERE.* A&A, 622, A156.
10. Zurlo, A., Cugno, G., Montesinos, M., Perez, S., Canovas, H., Casassus, S., Christiaens, V., Cieza, L., & Huelamo, N. (2020). *The widest Hα survey of accreting protoplanets around nearby transition disks.* A&A, 633, A119.
11. Wagner, K., Stone, J. M., Spalding, E., Apai, D., Dong, R., Ertel, S., Leisenring, J., & Webster, R. (2019). *Thermal Infrared Imaging of MWC 758 with the Large Binocular Telescope: Planetary-driven Spiral Arms?* ApJ, 882, 20.
12. Gaia Collaboration, Brown, A. G. A., Vallenari, A., Prusti, T., de Bruijne, J. H. J., Babusiaux, C., Biermann, M., Creevey, O. L., Evans, D. W., Eyer, L., Hutton, A., Jansen, F., Jordi, C., Klioner, S. A., Lammers, U., Lindegren, L., Luri, X., Mignard, F., Panem, C., Pourbaix, D., Randich, S., Sartoretti, P., Soubiran, C., Walton, N. A., Arenou, F., Bailer-Jones, C. A. L., Bastian, U., Cropper, M., Drimmel, R., Katz, D., Lattanzi, M. G., van Leeuwen, F., Bakker, J., Cacciari, C., Castañeda, J., De Angeli, F., Ducourant, C., Fabricius, C., Fouesneau, M., Frémat, Y., Guerra, R., Guerrier, A., Guiraud, J., Jean-Antoine Piccolo, A., Masana, E., Messineo, R., Mowlavi, N., Nicolas, C., Nienartowicz, K., Pailler, F., Panuzzo, P., Riclet, F., Roux, W., Seabroke, G. M., Sordo, R., Tanga, P., Thévenin, F., Gracia-Abril, G., Portell, J., Teyssier, D., Altmann, M., Andrae, R., Bellas-Velidis, I., Benson, K., Berthier, J., Blomme, R., Brugaletta, E., Burgess, P. W., Busso, G., Carry, B., Cellino, A., Cheek, N., Clementini, G., Damerdji, Y., Davidson, M., Delchambre, L., Dell'Oro, A., Fernández-Hernández, J., Galluccio, L., García-Lario, P., Garcia-Reinaldos, M., González-Núñez, J., Gosset, E., Haigron, R., Halbwachs, J.-L., Hambly, N. C., Harrison, D. L., Hatzidimitriou, D., Heiter, U., Hernández, J., Hestroffer, D., Hodgkin, S. T., Holl, B., Janßen, K., Jevardat de Fombelle, G., Jordan, S., Krone-Martins, A., Lanzafame, A. C., Löffler, W., Lorca, A., Manteiga, M., Marchal, O., Marrese, P. M., Moitinho, A., Mora, A., Muinonen, K., Osborne, P., Pancino, E., Pauwels, T., Petit, J.-M., Recio-Blanco, A., Richards, P. J., Riello, M., Rimoldini, L., Robin, A. C., Roegiers, T., Rybizki, J., Sarro, L. M., Siopis, C., Smith, M., Sozzetti, A., Ulla, A., Utrilla, E., van Leeuwen, M., van Reeven, W., Abbas, U., Abreu Aramburu, A., Accart, S., Aerts, C., Aguado, J. J., Ajaj, M., Altavilla, G., Álvarez, M. A., Álvarez Cid-Fuentes, J., Alves, J., Anderson, R. I., Anglada Varela, E., Antoja, T., Audard, M., Baines, D., Baker, S. G., Balaguer-Núñez, L., Balbinot, E., Balog, Z., Barache, C., Barbato, D., Barros, M., Barstow, M. A., Bartolomé, S., Bassilana, J.-L., Bauchet, N., Baudesson-Stella, A., Becciani, U., Bellazzini, M., Bernet, M., Bertone, S., Bianchi, L., Blanco-Cuaresma, S., Boch, T., Bombrun, A., Bossini, D., Bouquillon, S., Bragaglia, A., Bramante, L., Breedt, E., Bressan, A., Brouillet, N., Bucciarelli, B., Burlacu, A., Busonero, D., Butkevich, A. G., Buzzi, R., Caffau, E., Cancelliere, R., Cánovas, H., Cantat-Gaudin, T., Carballo, R., Carlucci, T., Carnerero, M. I., Carrasco, J. M., Casamiquela, L., Castellani, M., Castro-Ginard, A., Castro Sampol, P., Chaoul, L., Charlot, P., Chemin, L., Chiavassa, A., Cioni, M.-R. L., Comoretto, G., Cooper, W. J., Cornez, T., Cowell, S., Crifo, F., Crosta, M., Crowley, C., Dafonte, C., Dapergolas, A., David, M., David, P., de Laverny, P., De Luise, F., De March, R., De Ridder, J., de Souza, R., de Teodoro, P., de Torres, A., del Peloso, E. F., del Pozo, E., Delbo, M., Delgado, A., Delgado, H. E., Delisle, J.-B., Di Matteo, P., Diakite, S., Diener, C., Distefano, E., Dolding, C., Eappachen, D., Edvardsson, B., Enke, H., Esquej, P., Fabre, C., Fabrizio, M., Faigler, S., Fedorets, G., Fernique, P., Fienga, A., Figueras, F., Fouron, C., Fragkoudi, F., Fraile, E., Franke, F., Gai, M., Garabato, D., Garcia-Gutierrez, A., García-Torres, M., Garofalo, A., Gavras, P., Gerlach, E., Geyer, R., Giacobbe, P., Gilmore, G., Girona, S., Giuffrida, G., Gomel, R., Gomez, A., Gonzalez-Santamaria, I., González-Vidal, J. J., Granvik, M., Gutiérrez-Sánchez, R., Guy, L. P., Hauser, M., Haywood, M., Helmi, A., Hidalgo, S. L., Hilger, T., Hładczuk, N., Hobbs, D., Holland, G., Huckle, H. E., Jasniewicz, G., Jonker, P. G., Juaristi Campillo, J., Julbe, F., Karbevska, L., Kervella, P., Khanna, S., Kochoska, A., Kontizas, M., Kordopatis, G., Korn, A. J., Kostrzewa-Rutkowska, Z., Kruszyńska, K., Lambert, S., Lanza, A. F., Lasne, Y., Le Campion, J.-F., Le

Fustec, Y., Lebreton, Y., Lebzelter, T., Leccia, S., Leclerc, N., Lecoeur-Taibi, I., Liao, S., Licata, E., Lindstrøm, E. P., Lister, T. A., Livanou, E., Lobel, A., Madrero Pardo, P., Managau, S., Mann, R. G., Marchant, J. M., Marconi, M., Marcos Santos, M. M. S., Marinoni, S., Marocco, F., Marshall, D. J., Martin Polo, L., Martín-Fleitas, J. M., Masip, A., Massari, D., Mastrobuono-Battisti, A., Mazeh, T., McMillan, P. J., Messina, S., Michalik, D., Millar, N. R., Mints, A., Molina, D., Molinaro, R., Molnár, L., Montegriffo, P., Mor, R., Morbidelli, R., Morel, T., Morris, D., Mulone, A. F., Munoz, D., Muraveva, T., Murphy, C. P., Musella, I., Noval, L., Ordénovic, C., Orrù, G., Osinde, J., Pagani, C., Pagano, I., Palaversa, L., Palicio, P. A., Panahi, A., Pawlak, M., Peñalosa Esteller, X., Penttilä, A., Piersimoni, A. M., Pineau, F.-X., Plachy, E., Plum, G., Poggio, E., Poretti, E., Poujoulet, E., Prša, A., Pulone, L., Racero, E., Ragaini, S., Rainer, M., Raiteri, C. M., Rambaux, N., Ramos, P., Ramos-Lerate, M., Re Fiorentin, P., Regibo, S., Reylé, C., Ripepi, V., Riva, A., Rixon, G., Robichon, N., Robin, C., Roelens, M., Rohrbasser, L., Romero-Gómez, M., Rowell, N., Royer, F., Rybicki, K. A., Sadowski, G., Sagristà Sellés, A., Sahlmann, J., Salgado, J., Salguero, E., Samaras, N., Sanchez Gimenez, V., Sanna, N., Santoveña, R., Sarasso, M., Schultheis, M., Sciacca, E., Segol, M., Segovia, J. C., Ségransan, D., Semeux, D., Shahaf, S., Siddiqui, H. I., Siebert, A., Siltala, L., Slezak, E., Smart, R. L., Solano, E., Solitro, F., Souami, D., Souchay, J., Spagna, A., Spoto, F., Steele, I. A., Steidelmüller, H., Stephenson, C. A., Süveges, M., Szabados, L., Szegedi-Elek, E., Taris, F., Tauran, G., Taylor, M. B., Teixeira, R., Thuillot, W., Tonello, N., Torra, F., Torra, J., Turon, C., Unger, N., Vaillant, M., van Dillen, E., Vanel, O., Vecchiato, A., Viala, Y., Vicente, D., Voutsinas, S., Weiler, M., Wevers, T., Wyrzykowski, Ł., Yoldas, A., Yvard, P., Zhao, H., Zorec, J., Zucker, S., Zurbach, C., & Zwitter, T. (2021). *Gaia Early Data Release 3. Summary of the contents and survey properties.* A&A, 649, A1.
13. Meeus, G., Montesinos, B., Mendigutía, I., Kamp, I., Thi, W. F., Eiroa, C., Grady, C. A., Mathews, G., Sandell, G., Martin-Zaïdi, C., Brittain, S., Dent, W. R. F., Howard, C., Ménard, F., Pinte, C., Roberge, A., Vandenbussche, B., & Williams, J. P. (2012). *Observations of Herbig Ae/Be stars with Herschel/PACS. The atomic and molecular contents of their protoplanetary discs.* A&A, 544, A78.
14. Grady, C. A., Muto, T., Hashimoto, J., Fukagawa, M., Currie, T., Biller, B., Thalmann, C., Sitko, M. L., Russell, R., Wisniewski, J., Dong, R., Kwon, J., Sai, S., Hornbeck, J., Schneider, G., Hines, D., Moro Martín, A., Feldt, M., Henning, T., Pott, J.-U., Bonnefoy, M., Bouwman, J., Lacour, S., Mueller, A., Juhász, A., Crida, A., Chauvin, G., Andrews, S., Wilner, D., Kraus, A., Dahm, S., Robitaille, T., Jang-Condell, H., Abe, L., Akiyama, E., Brandner, W., Brandt, T., Carson, J., Egner, S., Follette, K. B., Goto, M., Guyon, O., Hayano, Y., Hayashi, M., Hayashi, S., Hodapp, K., Ishii, M., Iye, M., Janson, M., Kandori, R., Knapp, G., Kudo, T., Kusakabe, N., Kuzuhara, M., Mayama, S., McElwain, M., Matsuo, T., Miyama, S., Morino, J.-I., Nishimura, T., Pyo, T.-S., Serabyn, G., Suto, H., Suzuki, R., Takami, M., Takato, N., Terada, H., Tomono, D., Turner, E., Watanabe, M., Yamada, T., Takami, H., Usuda, T., & Tamura, M. (2013). *Spiral Arms in the Asymmetrically Illuminated Disk of MWC 758 and Constraints on Giant Planets.* ApJ, 762, 48.
15. Benisty, M., Juhasz, A., Boccaletti, A., Avenhaus, H., Milli, J., Thalmann, C., Dominik, C., Pinilla, P., Buenzli, E., Pohl, A., Beuzit, J.-L., Birnstiel, T., de Boer, J., Bonnefoy, M., Chauvin, G., Christiaens, V., Garufi, A., Grady, C., Henning, T., Huelamo, N., Isella, A., Langlois, M., Ménard, F., Mouillet, D., Olofsson, J., Pantin, E., Pinte, C., & Pueyo, L. (2015). *Asymmetric features in the protoplanetary disk MWC 758.* A&A, 578, L6.
16. Reggiani, M., Christiaens, V., Absil, O., Mawet, D., Huby, E., Choquet, E., Gomez Gonzalez, C. A., Ruane, G., Femenia, B., Serabyn, E., Matthews, K., Barraza, M., Carlomagno, B., Defrère, D., Delacroix, C., Habraken, S., Jolivet, A., Karlsson, M., Orban de Xivry, G., Piron, P., Surdej, J., Vargas Catalan, E., & Wertz, O. (2018). *Discovery of a point-like source and a third spiral arm in the transition disk around the Herbig Ae star MWC 758.* A&A, 611, A74.
17. Fung, J., & Dong, R. (2015). *Inferring Planet Mass from Spiral Structures in Protoplanetary Disks.* ApJL, 815, L21.
18. Boccaletti, A., Pantin, E., Ménard, F., Galicher, R., Langlois, M., Benisty, M., Gratton, R., Chauvin, G., Ginski, C., Lagrange, A.-M., Zurlo, A., Biller, B., Bonavita, M., Bonnefoy, M., Brown-Sevilla, S., Cantalloube, F., Desidera, S., D'Orazi, V., Feldt, M., Hagelberg, J., Lazzoni, C., Mesa, D., Meyer, M., Perrot,

C., Vigan, A., Sauvage, J.-F., Ramos, J., Rousset, G., & Magnard, Y. (2021). *Investigating point sources in MWC 758 with SPHERE*. A&A, 652, L8.
19. Yu, S.-Y., Ho, L. C., & Zhu, Z. (2019). *A Tight Relation between Spiral Arm Pitch Angle and Protoplanetary Disk Mass*. ApJ, 877, 100.
20. Kratter, K., & Lodato, G. (2016). *Gravitational Instabilities in Circumstellar Disks*. ARAA, 54, 271.
21. Ren, B., Dong, R., Esposito, T. M., Pueyo, L., Debes, J. H., Poteet, C. A., Choquet, É., Benisty, M., Chiang, E., Grady, C. A., Hines, D. C., Schneider, G., & Soummer, R. (2018). *A Decade of MWC 758 Disk Images: Where Are the Spiral-arm-driving Planets?*. ApJL, 857, L9.
22. Ren, B., Dong, R., van Holstein, R. G., Ruffio, J.-B., Calvin, B. A., Girard, J. H., Benisty, M., Boccaletti, A., Esposito, T. M., Choquet, É., Mawet, D., Pueyo, L., Stolker, T., Chiang, E., Boer, J., Debes, J. H., Garufi, A., Grady, C. A., Hines, D. C., Maire, A.-L., Ménard, F., Millar-Blanchaer, M. A., Perrin, M. D., Poteet, C. A., & Schneider, G. (2020). *Dynamical Evidence of a Spiral Arm-driving Planet in the MWC 758 Protoplanetary Disk*. ApJL, 898, L38.
23. Hinz, P. M., Defrère, D., Skemer, A., Bailey, V., Stone, J., Spalding, E., Vaz, A., Pinna, E., Puglisi, A., Esposito, S., Montoya, M., Downey, E., Leisenring, J., Durney, O., Hoffmann, W., Hill, J., Millan-Gabet, R., Mennesson, B., Danchi, W., Morzinski, K., Grenz, P., Skrutskie, M., & Ertel, S. (2016). *Overview of LBTI: a multipurpose facility for high spatial resolution observations*. ProcSPIE, 9907, 990704.
24. Skrutskie, M. F., Jones, T., Hinz, P., Garnavich, P., Wilson, J., Nelson, M., Solheid, E., Durney, O., Hoffmann, W., Vaitheeswaran, V., McMahon, T., Leisenring, J., & Wong, A. (2010). *The Large Binocular Telescope mid-infrared camera (LMIRcam): final design and status*. Proc. SPIE, 7735, 77353H.
25. Leisenring, J. M., Skrutskie, M. F., Hinz, P. M., Skemer, A., Bailey, V., Eisner, J., Garnavich, P., Hoffmann, W. F., Jones, T., Kenworthy, M., Kuzmenko, P., Meyer, M., Nelson, M., Rodigas, T. J., Wilson, J. C., & Vaitheeswaran, V. (2012). *On-sky operations and performance of LMIRcam at the Large Binocular Telescope*. Proc. SPIE, 8446, 84464F.
26. Skemer, A. J., Hinz, P., Stone, J., Skrutskie, M., Woodward, C. E., Leisenring, J., & Briesemeister, Z. (2018). *ALES: overview and upgrades*. ProcSPIE, 10702, 107020C.
27. Stone, J. M., Skemer, A. J., Hinz, P., Briesemeister, Z., Barman, T., Woodward, C. E., Skrutskie, M., & Leisenring, J. (2018). *On-sky operations with the ALES integral field spectrograph*. ProcSPIE, 10702, 107023F.
28. Rayner, J. T., Toomey, D. W., Onaka, P. M., Denault, A. J., Stahlberger, W. E., Vacca, W. D., Cushing, M. C., & Wang, S. (2003). *SpeX: A Medium-Resolution 0.8-5.5 Micron Spectrograph and Imager for the NASA Infrared Telescope Facility*. PASP, 115, 362.
29. Cushing, M. C., Rayner, J. T., & Vacca, W. D. (2005). *An Infrared Spectroscopic Sequence of M, L, and T Dwarfs*. ApJ, 623, 1115.
30. Karalidi, T., Marley, M., Fortney, J. J., Morley, C., Saumon, D., Lupu, R., Visscher, C., & Freedman, R. (2021). *The Sonora Substellar Atmosphere Models. II. Cholla: A Grid of Cloud-free, Solar Metallicity Models in Chemical Disequilibrium for the JWST Era*. ApJ, 923, 269.
31. Beuzit, J.-L., Vigan, A., Mouillet, D., Dohlen, K., Gratton, R., Boccaletti, A., Sauvage, J.-F., Schmid, H. M., Langlois, M., Petit, C., Baruffolo, A., Feldt, M., Milli, J., Wahhaj, Z., Abe, L., Anselmi, U., Antichi, J., Barette, R., Baudrand, J., Baudoz, P., Bazzon, A., Bernardi, P., Blanchard, P., Brast, R., Bruno, P., Buey, T., Carbillet, M., Carle, M., Cascone, E., Chapron, F., Charton, J., Chauvin, G., Claudi, R., Costille, A., De Caprio, V., de Boer, J., Delboulbé, A., Desidera, S., Dominik, C., Downing, M., Dupuis, O., Fabron, C., Fantinel, D., Farisato, G., Feautrier, P., Fedrigo, E., Fusco, T., Gigan, P., Ginski, C., Girard, J., Giro, E., Gisler, D., Gluck, L., Gry, C., Henning, T., Hubin, N., Hugot, E., Incorvaia, S., Jaquet, M., Kasper, M., Lagadec, E., Lagrange, A.-M., Le Coroller, H., Le Mignant, D., Le Ruyet, B., Lessio, G., Lizon, J.-L., Llored, M., Lundin, L., Madec, F., Magnard, Y., Marteaud, M., Martinez, P., Maurel, D., Ménard, F., Mesa, D., Möller-Nilsson, O., Moulin, T., Moutou, C., Origné, A., Parisot, J., Pavlov, A., Perret, D., Pragt, J., Puget, P., Rabou, P., Ramos, J., Reess, J.-M., Rigal, F., Rochat, S., Roelfsema, R., Rousset, G., Roux, A., Saisse, M., Salasnich, B., Santambrogio, E., Scuderi, S., Segransan, D., Sevin, A., Siebenmorgen, R., Soenke, C., Stadler, E., Suarez, M., Tiphène, D., Turatto, M., Udry, S., Vakili, F., Waters, L. B. F. M., Weber, L., Wildi, F., Zins, G., & Zurlo, A. (2019). *SPHERE: the exoplanet imager for the Very Large Telescope*. A&A, 631, A155.

32. Dong, R., Liu, S.-. yuan ., Eisner, J., Andrews, S., Fung, J., Zhu, Z., Chiang, E., Hashimoto, J., Liu, H. B., Casassus, S., Esposito, T., Hasegawa, Y., Muto, T., Pavlyuchenkov, Y., Wilner, D., Akiyama, E., Tamura, M., & Wisniewski, J. (2018). *The Eccentric Cavity, Triple Rings, Two-armed Spirals, and Double Clumps of the MWC 758 Disk*. ApJ, 860, 124.
33. Fung, J., Shi, J.-M., & Chiang, E. (2014). How Empty are Disk Gaps Opened by Giant Planets? ApJ, 782, 88.
34. Allard, F., Homeier, D., Freytag, B. (2012). *Models of very-low-mass stars, brown dwarfs, and exoplanets.* Philosophical Transactions of the Royal Society, 370, 2765.
35. Marley, M. S., Fortney, J. J., Hubickyj, O., Bodenheimer, P., & Lissauer, J. J. (2007). *On the Luminosity of Young Jupiters*. ApJ, 655, 541.
36. Wang, S., & Chen, X. (2019). *The Optical to Mid-infrared Extinction Law Based on the APOGEE, Gaia DR2, Pan-STARRS1, SDSS, APASS, 2MASS, and WISE Surveys*. ApJ, 877, 116.
37. Szulágyi, J., Dullemond, C. P., Pohl, A., & Quanz, S. P. (2019). *Observability of forming planets and their circumplanetary discs II. - SEDs and near-infrared fluxes*. MNRAS, 487, 1248.
38. Sanchis, E., Picogna, G., Ercolano, B., Testi, L., & Rosotti, G. (2020). *Detectability of embedded protoplanets from hydrodynamical simulations*. MNRAS, 492, 3440.
39. Chen, X., & Szulágyi, J. (2021). *Observability of Forming Planets and their Circumplanetary Disks IV. -- with JWST & ELT*. arXiv e-prints, arXiv:2112.12821.
40. Spiegel, D. S., & Burrows, A. (2012). *Spectral and Photometric Diagnostics of Giant Planet Formation Scenarios.* ApJ, 745, 174.
41. Marley, M. S., Saumon, D., Visscher, C., Lupu, R., Freedman, R., Morley, C., Fortney, J. J., Seay, C., Smith, A. J. R. W., Teal, D. J., & Wang, R. (2021). *The Sonora Brown Dwarf Atmosphere and Evolution Models. I. Model Description and Application to Cloudless Atmospheres in Rainout Chemical Equilibrium*. ApJ, 920, 85.
42. Baraffe, I., Chabrier, G., Barman, T. S., Allard, F., & Hauschildt, P. H. (2003). *Evolutionary models for cool brown dwarfs and extrasolar giant planets. The case of HD 209458*. A&A, 402, 701.
43. Cugno, G., Patapis, P., Stolker, T., Quanz, S. P., Boehle, A., Hoeijmakers, H. J., Marleau, G.-D., Mollière, P., Nasedkin, E., & Snellen, I. A. G. (2021). *Molecular mapping of the PDS70 system. No molecular absorption signatures from the forming planet PDS70 b*. A&A, 653, A12.
44. Hashimoto, J., Aoyama, Y., Konishi, M., Uyama, T., Takasao, S., Ikoma, M., & Tanigawa, T. (2020). *Accretion Properties of PDS 70b with MUSE*. AJ, 159, 222.
45. Uyama, T., Xie, C., Aoyama, Y., Beichman, C. A., Hashimoto, J., Dong, R., Hasegawa, Y., Ikoma, M., Mawet, D., McElwain, M. W., Ruffio, J.-B., Wagner, K. R., Wang, J. J., & Zhou, Y. (2021). *Keck/OSIRIS Paβ High-contrast Imaging and Updated Constraints on PDS 70b*. AJ, 162, 214.
46. Wang, J. J., Vigan, A., Lacour, S., Nowak, M., Stolker, T., De Rosa, R. J., Ginzburg, S., Gao, P., Abuter, R., Amorim, A., Asensio-Torres, R., Bauböck, M., Benisty, M., Berger, J. P., Beust, H., Beuzit, J.-L., Blunt, S., Boccaletti, A., Bohn, A., Bonnefoy, M., Bonnet, H., Brandner, W., Cantalloube, F., Caselli, P., Charnay, B., Chauvin, G., Choquet, E., Christiaens, V., Clénet, Y., Coudé Du Foresto, V., Cridland, A., de Zeeuw, P. T., Dembet, R., Dexter, J., Drescher, A., Duvert, G., Eckart, A., Eisenhauer, F., Facchini, S., Gao, F., Garcia, P., Garcia Lopez, R., Gardner, T., Gendron, E., Genzel, R., Gillessen, S., Girard, J., Haubois, X., Heißel, G., Henning, T., Hinkley, S., Hippler, S., Horrobin, M., Houllé, M., Hubert, Z., Jiménez-Rosales, A., Jocou, L., Kammerer, J., Keppler, M., Kervella, P., Meyer, M., Kreidberg, L., Lagrange, A.-M., Lapeyrère, V., Le Bouquin, J.-B., Léna, P., Lutz, D., Maire, A.-L., Ménard, F., Mérand, A., Mollière, P., Monnier, J. D., Mouillet, D., Müller, A., Nasedkin, E., Ott, T., Otten, G. P. P. L., Paladini, C., Paumard, T., Perraut, K., Perrin, G., Pfuhl, O., Pueyo, L., Rameau, J., Rodet, L., Rodríguez-Coira, G., Rousset, G., Scheithauer, S., Shangguan, J., Shimizu, T., Stadler, J., Straub, O., Straubmeier, C., Sturm, E., Tacconi, L. J., van Dishoeck, E. F., Vincent, F., von Fellenberg, S. D., Ward-Duong, K., Widmann, F., Wieprecht, E., Wiezorrek, E., Woillez, J., & Gravity Collaboration (2021). *Constraining the Nature of the PDS 70 Protoplanets with VLTI/GRAVITY*. AJ, 161, 148.
47. Currie, T., Lawson, K., Schneider, G., Lyra, W., Wisniewski, J., Grady, C., Guyon, O., Tamura, M., Kotani, T., Kawahara, H., Brandt, T., Uyama, T., Muto, T., Dong, R., Kudo, T., Hashimoto, J., Fukagawa, M., Wagner, K., Lozi, J., Chilcote, J., Tobin, T., Groff, T., Ward-Duong, K., Januszewski, W., Norris, B., Tuthill,

P., van der Marel, N., Sitko, M., Deo, V., Vievard, S., Jovanovic, N., Martinache, F., & Skaf, N. (2022). *Images of embedded Jovian planet formation at a wide separation around AB Aurigae*. Nature Astronomy, 6, 751.

48. Fung, J., Zhu, Z., & Chiang, E. (2019). *Circumplanetary Disk Dynamics in the Isothermal and Adiabatic Limits*. ApJ, 887, 152.
49. Shuai, L., Ren, B. B., Dong, R., Zhou, X., Pueyo, L., De Rosa, R. J., Fang, T., & Mawet, D. (2022). *Stellar Flyby Analysis for Spiral Arm Hosts with Gaia DR3*. arXiv e-prints, arXiv:2210.03725.
50. Zapata, L. A., Rodríguez, L. F., Fernández-López, M., Palau, A., Estalella, R., Osorio, M., Anglada, G., & Huelamo, N. (2020). *Tidal Interaction between the UX Tauri A/C Disk System Revealed by ALMA*. ApJ, 896, 132.
51. Kuo, I.-H. G., Yen, H.-W., Gu, P.-G., & Chang, T.-E. (2022). *Kinematical Constraint on Eccentricity in the Protoplanetary Disk MWC 758 with ALMA*. ApJ, 938, 50.
52. Zhu, Z., Dong, R., Stone, J. M., & Rafikov, R. R. (2015). *The Structure of Spiral Shocks Excited by Planetary-mass Companions.* ApJ, 813, 88.
53. Boehler, Y., Ricci, L., Weaver, E., Isella, A., Benisty, M., Carpenter, J., Grady, C., Shen, B.-T., Tang, Y.-W., & Perez, L. (2018). T*he Complex Morphology of the Young Disk MWC 758: Spirals and Dust Clumps around a Large Cavity*. ApJ, 853, 162.
54. Li, H., Finn, J. M., Lovelace, R. V. E., & Colgate, S. A. (2000). *Rossby Wave Instability of Thin Accretion Disks. II. Detailed Linear Theory*. ApJ, 533, 1023.
55. Hammer, M., Pinilla, P., Kratter, K. M., & Lin, M.-K. (2019). *Observational diagnostics of elongated planet-induced vortices with realistic planet formation time-scales*. MNRAS, 482, 3609.
56. Morley, C. V., Skemer, A. J., Allers, K. N., Marley, M. S., Faherty, J. K., Visscher, C., Beiler, S. A., Miles, B. E., Lupu, R., Freedman, R. S., Fortney, J. J., Geballe, T. R., & Bjoraker, G. L. (2018). *An L Band Spectrum of the Coldest Brown Dwarf*. ApJ, 858, 97.
57. Macintosh, B., Graham, J. R., Barman, T., De Rosa, R. J., Konopacky, Q., Marley, M. S., Marois, C., Nielsen, E. L., Pueyo, L., Rajan, A., Rameau, J., Saumon, D., Wang, J. J., Patience, J., Ammons, M., Arriaga, P., Artigau, E., Beckwith, S., Brewster, J., Bruzzone, S., Bulger, J., Burningham, B., Burrows, A. S., Chen, C., Chiang, E., Chilcote, J. K., Dawson, R. I., Dong, R., Doyon, R., Draper, Z. H., Duchêne, G., Esposito, T. M., Fabrycky, D., Fitzgerald, M. P., Follette, K. B., Fortney, J. J., Gerard, B., Goodsell, S., Greenbaum, A. Z., Hibon, P., Hinkley, S., Cotten, T. H., Hung, L.-W., Ingraham, P., Johnson-Groh, M., Kalas, P., Lafreniere, D., Larkin, J. E., Lee, J., Line, M., Long, D., Maire, J., Marchis, F., Matthews, B. C., Max, C. E., Metchev, S., Millar-Blanchaer, M. A., Mittal, T., Morley, C. V., Morzinski, K. M., Murray-Clay, R., Oppenheimer, R., Palmer, D. W., Patel, R., Perrin, M. D., Poyneer, L. A., Rafikov, R. R., Rantakyrö, F. T., Rice, E. L., Rojo, P., Rudy, A. R., Ruffio, J.-B., Ruiz, M. T., Sadakuni, N., Saddlemyer, L., Salama, M., Savransky, D., Schneider, A. C., Sivaramakrishnan, A., Song, I., Soummer, R., Thomas, S., Vasisht, G., Wallace, J. K., Ward-Duong, K., Wiktorowicz, S. J., Wolff, S. G., & Zuckerman, B. (2015). *Discovery and spectroscopy of the young jovian planet 51 Eri b with the Gemini Planet Imager*. Science, 350, 64.
58. Skemer, A. J., Morley, C. V., Allers, K. N., Geballe, T. R., Marley, M. S., Fortney, J. J., Faherty, J. K., Bjoraker, G. L., & Lupu, R. (2016). *The First Spectrum of the Coldest Brown Dwarf*.
59. Follette, K. B., Close, L. M., Males, J. R., Ward-Duong, K., Balmer, W. O., Adams Redai, J., Morales, J., Sarosi, C., Dacus, B., De Rosa, R. J., Garcia Toro, F., Leonard, C., Macintosh, B., Morzinski, K. M., Mullen, W., Palmo, J., Nzaba Saitoti, R., Spiro, E., Treiber, H., Wang, J., Wang, D., Watson, A., & Weinberger, A. J. (2022). *The Giant Accreting Protoplanet Survey (GAPlanetS) -- Results from a Six Year Campaign to Image Accreting Protoplanets.* arXiv e-prints, arXiv:2211.02109.
60. Briesemeister, Z., Skemer, A. J., Stone, J. M., Stelter, R. D., Hinz, P., Leisenring, J., Skrutskie, M. F., Woodward, C. E., & Barman, T. (2018). *MEAD: data reduction pipeline for ALES integral field spectrograph and LBTI thermal infrared calibration unit*. Proc. SPIE, 10702, 107022Q.
61. Stone, J. M., Barman, T., Skemer, A. J., Briesemeister, Z. W., Brock, L. S., Hinz, P. M., Leisenring, J. M., Woodward, C. E., Skrutskie, M. F., & Spalding, E. (2020). *High-contrast Thermal Infrared Spectroscopy with ALES: The 3-4 µm Spectrum of κ Andromedae b*. AJ, 160, 262.

62. Doelman, D. S., Stone, J. M., Briesemeister, Z. W., Skemer, A. J. I., Barman, T., Brock, L. S., Hinz, P. M., Bohn, A., Kenworthy, M., Haffert, S. Y., Snik, F., Ertel, S., Leisenring, J. M., Woodward, C. E., & Skrutskie, M. F. (2022). *L-band Integral Field Spectroscopy of the HR 8799 Planetary System*. AJ, 163, 217.
63. Marois, C., Lafrenière, D., Doyon, R., Macintosh, B., & Nadeau, D. (2006). *Angular Differential Imaging: A Powerful High-Contrast Imaging Technique*. ApJ, 641, 556.
64. Soummer, R., Pueyo, L., & Larkin, J. (2012). *Detection and Characterization of Exoplanets and Disks Using Projections on Karhunen-Loève Eigenimages*. ApJL, 755, L28.
65. Bottom, M., Ruane, G., & Mawet, D. (2017). *Noise-weighted Angular Differential Imaging*. Research Notes of the American Astronomical Society, 1, 30.
66. Wagner, K., Apai, D., Kasper, M., McClure, M., Robberto, M., & Currie, T. (2020). *Direct Imaging Discovery of a Young Brown Dwarf Companion to an A2V Star*. ApJL, 902, L6.
67. Cushing, M. C., Vacca, W. D., & Rayner, J. T. (2004). *Spextool: A Spectral Extraction Package for SpeX, a 0.8-5.5 Micron Cross-Dispersed Spectrograph*. PASP, 116, 362.
68. Lagrange, A.-M., Bonnefoy, M., Chauvin, G., Apai, D., Ehrenreich, D., Boccaletti, A., Gratadour, D., Rouan, D., Mouillet, D., Lacour, S., & Kasper, M. (2010). *A Giant Planet Imaged in the Disk of the Young Star β Pictoris*. Science, 329, 57.
69. Greco, J. P., & Brandt, T. D. (2016). *The Measurement, Treatment, and Impact of Spectral Covariance and Bayesian Priors in Integral-field Spectroscopy of Exoplanets*. ApJ, 833, 134.
70. Chabrier, G., Baraffe, I., Allard, F., & Hauschildt, P. (2000). *Evolutionary Models for Very Low-Mass Stars and Brown Dwarfs with Dusty Atmospheres*. ApJ, 542, 464.
71. Miles, B. E., Skemer, A. J. I., Morley, C. V., Marley, M. S., Fortney, J. J., Allers, K. N., Faherty, J. K., Geballe, T. R., Visscher, C., Schneider, A. C., Lupu, R., Freedman, R. S., & Bjoraker, G. L. (2020). *Observations of Disequilibrium CO Chemistry in the Coldest Brown Dwarfs*. AJ, 160, 63.
72. Stone, J. M., Skemer, A. J., Hinz, P. M., Bonavita, M., Kratter, K. M., Maire, A.-L., Defrere, D., Bailey, V. P., Spalding, E., Leisenring, J. M., Desidera, S., Bonnefoy, M., Biller, B., Woodward, C. E., Henning, T., Skrutskie, M. F., Eisner, J. A., Crepp, J. R., Patience, J., Weigelt, G., De Rosa, R. J., Schlieder, J., Brandner, W., Apai, D., Su, K., Ertel, S., Ward-Duong, K., Morzinski, K. M., Schertl, D., Hofmann, K.-H., Close, L. M., Brems, S. S., Fortney, J. J., Oza, A., Buenzli, E., & Bass, B. (2018). *The LEECH Exoplanet Imaging Survey: Limits on Planet Occurrence Rates under Conservative Assumptions*. AJ, 156, 286.

This document includes:
Supplementary Figures 1-11

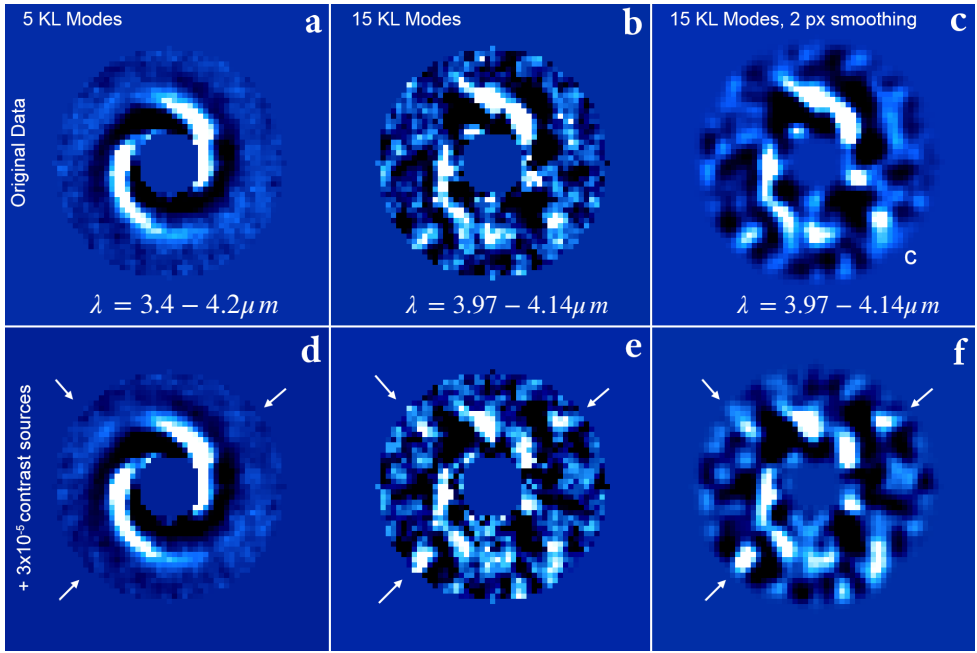

**Supplementary Figure 1.** Various reductions of the 20191114 LBTI/ALES data. The panels on the top (in particular **a**, **c**) are the versions shown in Fig. 1, whereas panels on the bottom (**d-f**) have three sources injected at -90°,+90°, and 180° from MWC 758c that are marked by arrows. The non-aggressive reduction (with 5 KL modes, panels **a,d**) does not show the injected sources, or MWC 758c, but does reveal the disk with less signal degradation compared to the reduction with 15 KL modes. Panel **b** illustrates that MWC 758c is detected at the pixel-to-pixel level.

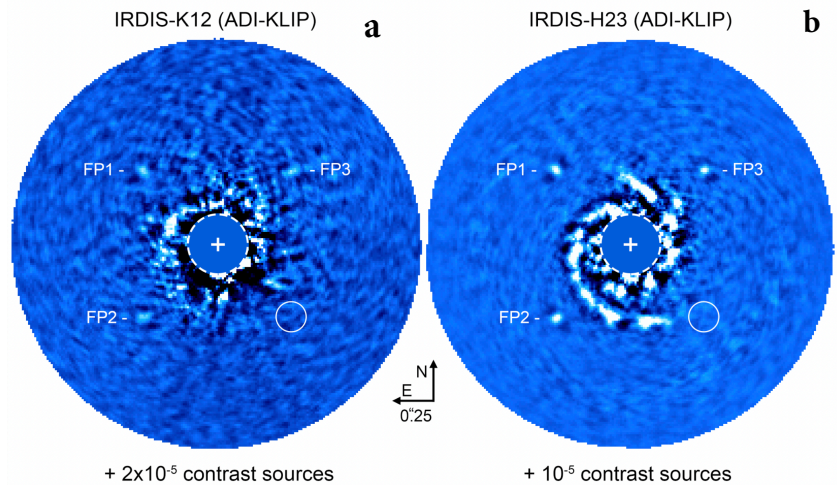

**Supplementary Figure S2.** SPHERE-IRDIS Images of MWC 758. Images taken in the *K12* (**a**; UT 2016-01-01) and *H23* (**b**; UT 2018-12-17) filters. Synthetic sources (fake planets, or FPs) were injected into the datasets to assess sensitivity limits. The position of MWC 758c is marked by a circle with radius equivalent to the FWHM of the PSF from LBTI/ALES.

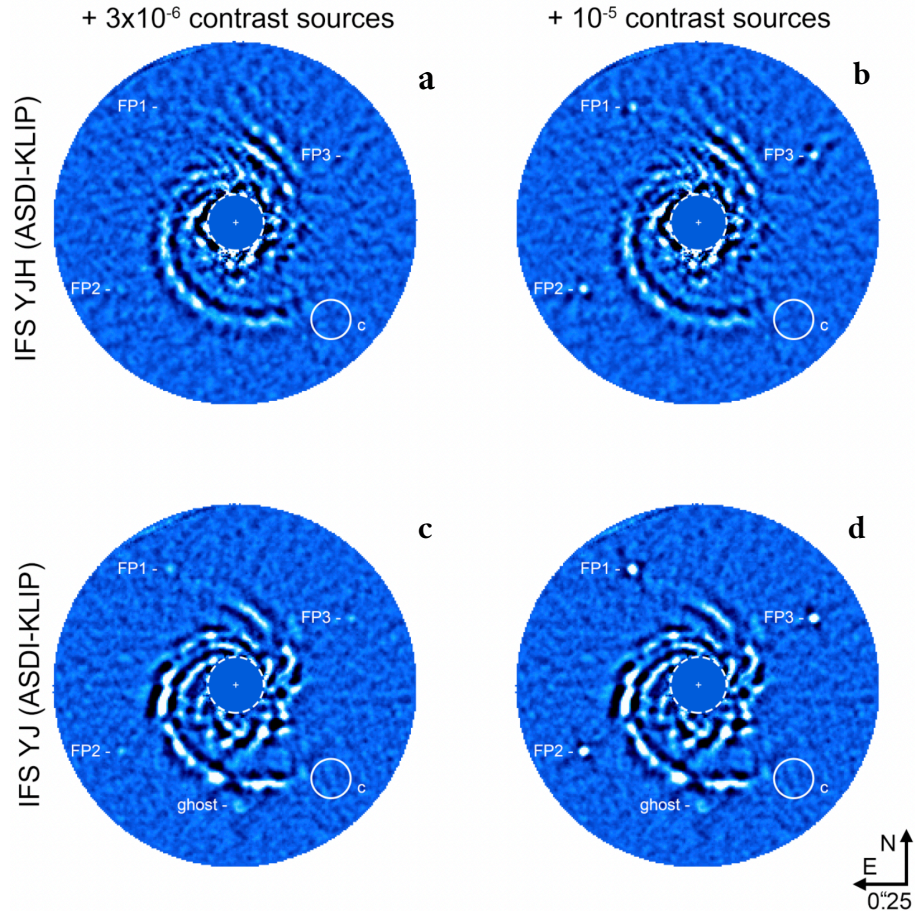

**Supplementary Figure 3.** SPHERE-IFS images of MWC 758. Data were taken in *YJH* (**a, b**; UT 2016-01-01) and *YJ* modes (**c, d**; UT 2018-12-17). Synthetic point sources (fake planets, or FPs) were injected into each dataset in order to assess detection limits.

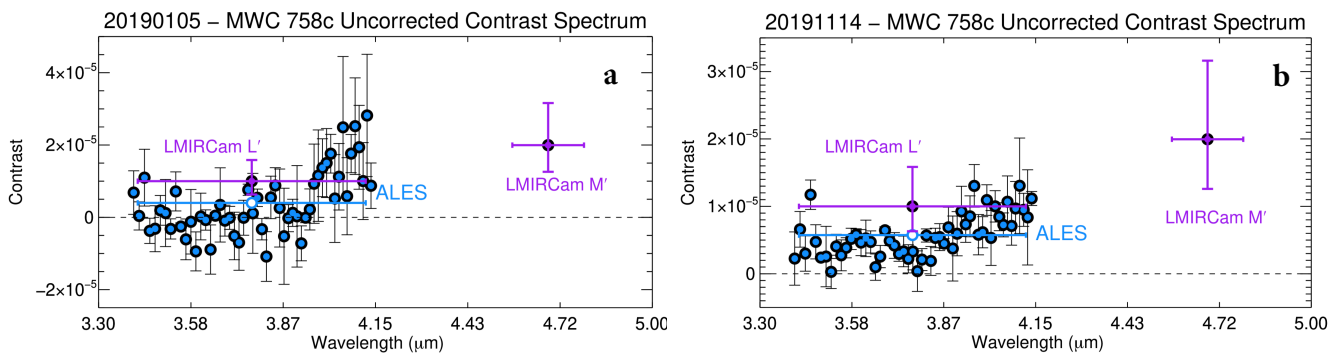

**Supplementary Figure 4.** Uncorrected ALES contrast spectra of MWC 758c from 20190105 (**a**) and 20191114 (**b**). The LMIRCam $L'$ and $M'$ photometry, reproduced from *11*, are corrected via negative synthetic planet injections (see step 3 below). The binned ALES $L'$ measurements (prior to correction) are consistent with the corrected $L'$ photometry at the $1\sigma$ level. Error bars represent the standard deviation of recovered flux of three injected sources.

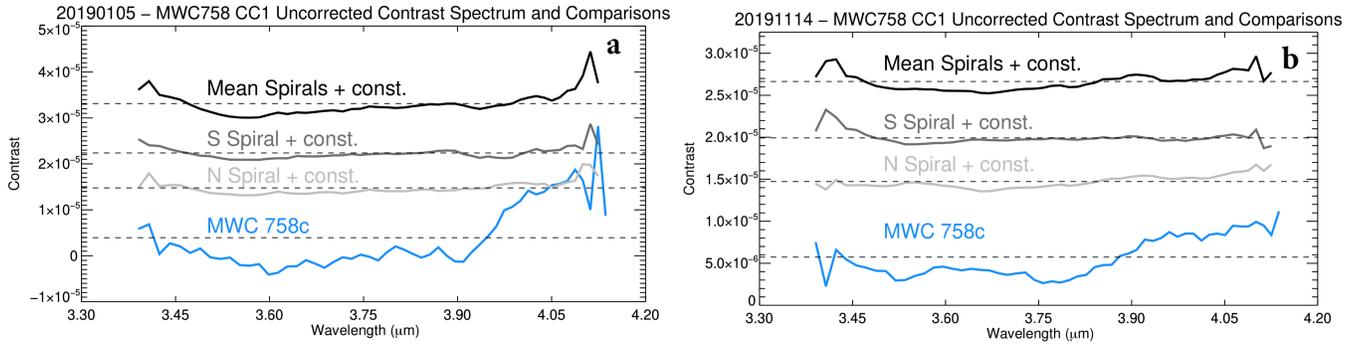

**Supplementary Figure 5.** Comparisons between extracted spectra of MWC 758c and the circumstellar disk from 20190105 (**a**) and 20191114 (**b**). The flatness of the disk's contrast spectrum is consistent with scattered light. Note that telluric absorption is strongest at the first and last few wavelength channels, causing sharp increases at these wavelengths.

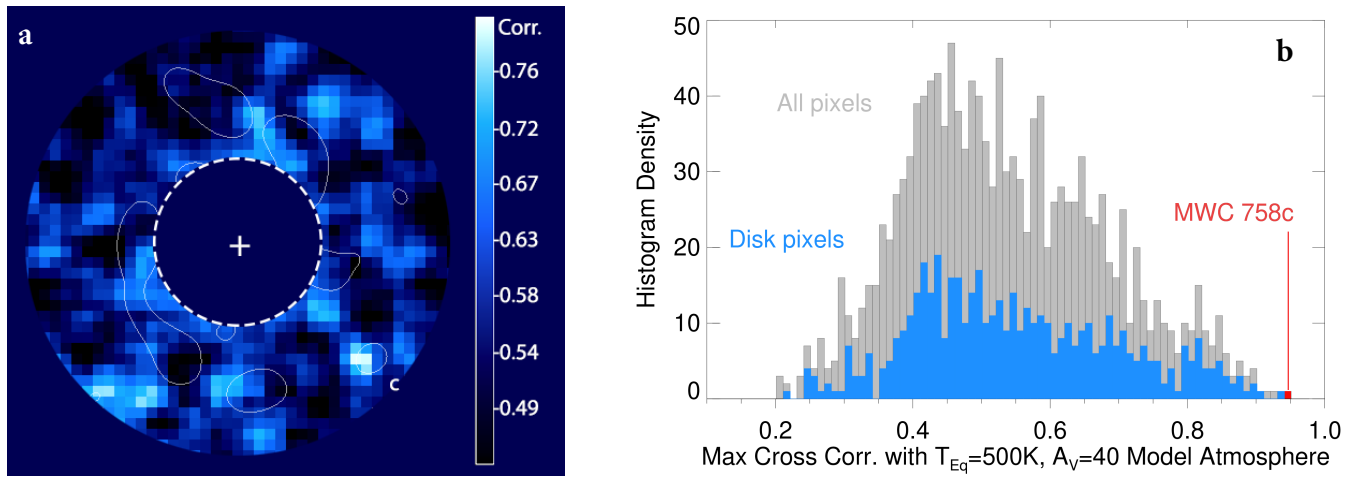

**Supplementary Figure 6.** Cross correlation map and resulting histogram. Two-pixel smoothed cross-correlation map of each pixel's spectrum vs. a $T_{\text{eff}} = 500$ K, $A_V = 40$ BT-Settl spectrum (*32*) for the higher-quality 20191114 dataset (**a**) and histogram (**b**, pre-smoothing), which demonstrates that MWC 758c stands out as the reddest point in the image. The contours show the disk and MWC 758c (i.e., the total intensity image).

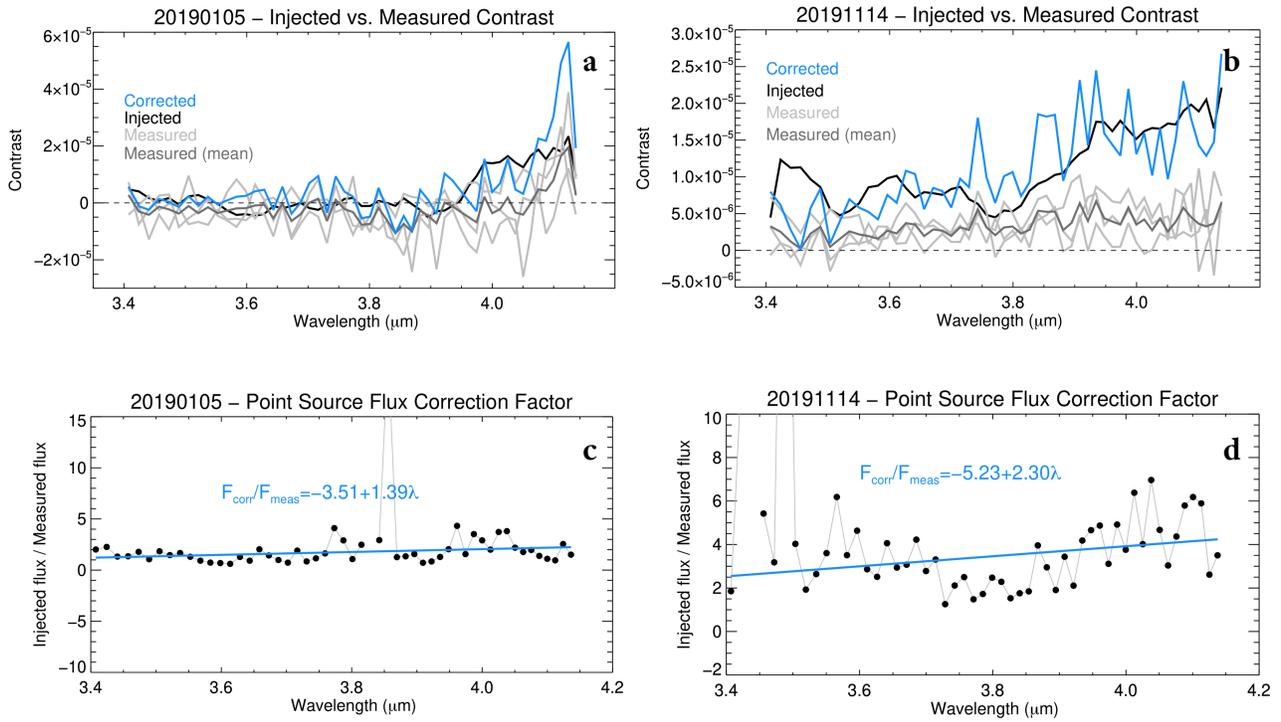

**Supplementary Figure 7.** Forward modeling results. Injected vs. measured contrast of injected sources (**a, b**) and derived point source flux correction models (**c, d**) from UT 2019-01-05 and 2019-11-14, respectively.

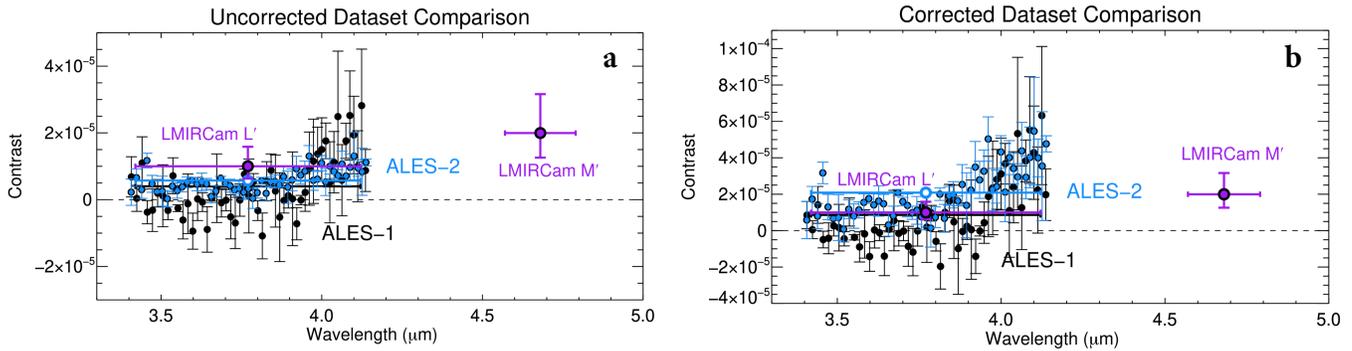

**Supplementary Figure 8.** Corrected contrast spectra of MWC 758c (**a**). The binned ALES $L'$ measurements (post-correction, **b**) are consistent with the corrected $L'$ photometry at the $\sim(0, 1)\sigma$ level (i.e., bracketing the previous LMIRCam measurement). Error bars represent the standard deviation of recovered flux of three injected sources.

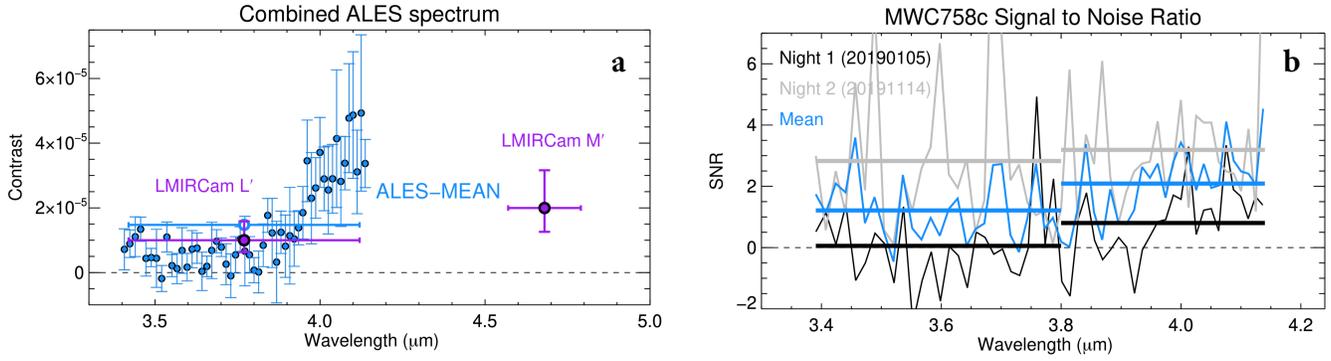

**Supplementary Figure 9.** Averaged ALES spectrum (**a**), which is consistent with the $L'$ photometry to <1$\sigma$. **b** Signal to noise ratio of the two ALES epochs and their averaged spectrum (with propagated uncertainties). Error bars represent the standard deviation of recovered flux of three injected sources.

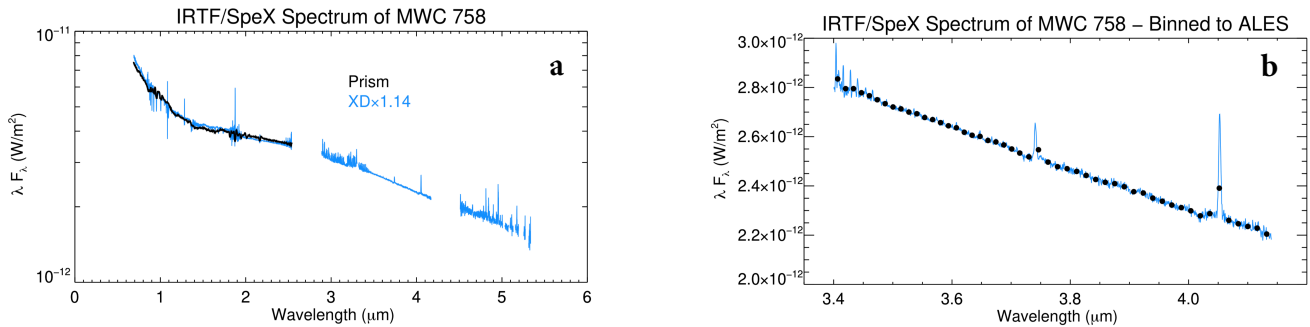

**Supplementary Figure 10. a** IRTF/SpeX spectrum of MWC 758 (including contributions from both the star and unresolved inner disk emission). The cross-dispersed (XD, blue) spectrum is scaled to the lower resolution prism spectrum (black) by 15% to account for light losses along the slit. **b** IRTF/SpeX spectrum of MWC 758 (blue) binned to the resolution of ALES (black).

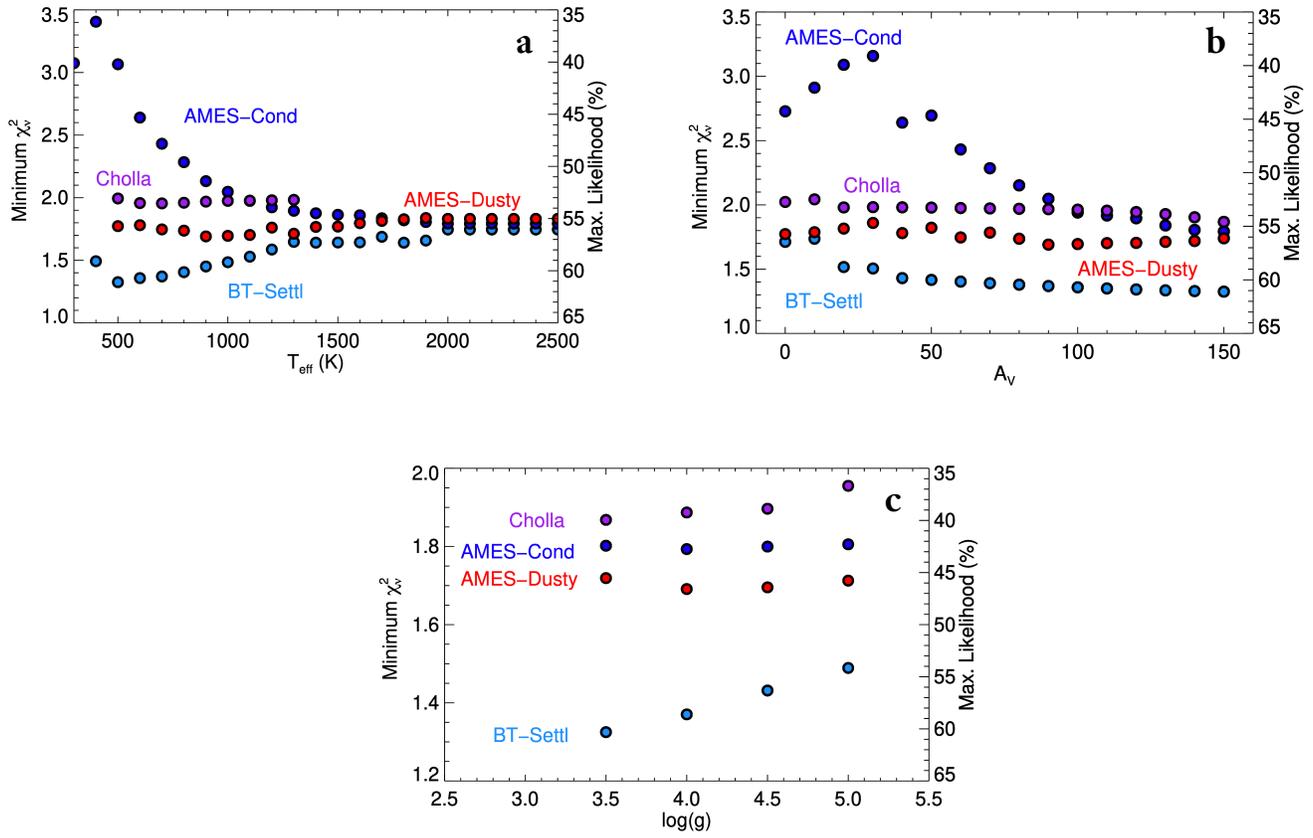

**Supplementary Figure 11.** Minimum reduced chi-squared ($\chi_\nu^2$) values and corresponding maximum likelihoods (**a** vs. effective temperature; **b** vs. $A_V$, **c** vs. surface gravity) for Cholla, COND, DUSTY, and BT-Settl models compared to the ALES and LMIRCam data.